\documentclass[lettersize,journal]{IEEEtran}
\usepackage[utf8]{inputenc}
\usepackage{amsmath,amssymb,amsfonts}

\usepackage{graphicx}
\usepackage[labelsep=period]{caption}
\usepackage[caption=false,font=normalsize,labelfont=sf,textfont=sf]{subfig}
\usepackage{subcaption}
\usepackage{stfloats}
\usepackage{tikz}
\newcommand{\circled}[2][]{%
  \tikz[baseline=(char.base)]{%
    \node[shape = circle, draw, inner sep = 0.31pt, minimum size = 1.15em]
    (char) {\phantom{\ifblank{#1}{#2}{#1}}};%
    \node at (char.center) {\makebox[0pt][c]{#2}};}}

\usepackage{multirow}
\usepackage{makecell}

\usepackage[ruled,vlined,linesnumbered,longend,resetcount]{algorithm2e}
\usepackage{algpseudocode}

\usepackage{textcomp}
\usepackage{xcolor}
\usepackage{soul}
\usepackage{enumitem}

\usepackage{cite}
\usepackage{url}
\usepackage{verbatim}
\usepackage{orcidlink}
\usepackage{hyperref}

\usepackage{xcolor}
\usepackage[normalem]{ulem}
\usepackage{soul}
\usepackage{tikz}

\usepackage{xcolor}

\definecolor{greenn}{HTML}{1C9F96}
\definecolor{pinkk}{HTML}{CC00FF}

\newcommand\reduline{%
  \bgroup\markoverwith{\textcolor{red}{\rule[0.5ex]{0.8pt}{1.5pt}}}%
  \ULon}

\newcommand\redst[1]{\textcolor{red}{\st{#1}}}

\begin{document}

\title{Curated Collaborative AI Edge with Network Data Analytics for B5G/6G Radio Access Networks}


\author{
    \IEEEauthorblockN{Sardar Jaffar Ali\orcidlink{0009-0002-5485-7598}\textsuperscript{1}, Syed M. Raza\orcidlink{0000-0001-6580-3232}\textsuperscript{2}, Duc-Tai Le\orcidlink{0000-0002-5286-6629}
    \textsuperscript{3}, 
    Min Young Chung\orcidlink{0000-0002-9063-8045}\textsuperscript{3}, Ness Shroff\orcidlink{0000-0002-4606-6879} \textsuperscript{4}, Hyunseung Choo\orcidlink{0000-0002-6485-3155}\textsuperscript{3,*}} \\
    \IEEEauthorblockA{
        \textsuperscript{1}Convergence Research Institute, Sungkyunkwan University, Suwon, South Korea \\
        \textsuperscript{2}James Watt School of Engineering, University of Glasgow, Glasgow, United Kingdom \\
        \textsuperscript{3}Dept. of Electrical and Computer Engineering, Sungkyunkwan University, Suwon, South Korea \\
        \textsuperscript{4}Department of Electrical and Computer Engineering, Ohio State University, Columbus, USA \\
        \textsuperscript{*}Corresponding author (choo@skku.edu) \\
    }
}
\markboth{Journal of \LaTeX\ Class Files,~Vol.~15, No.~19, September~2021}%
{Shell \MakeLowercase{\textit{et al.}}: A Sample Article Using IEEEtran.cls for IEEE Journals}


\maketitle

\begin{abstract}

Despite advancements, Radio Access Networks (RAN) still account for over 50\% of the total power consumption in 5G networks. Existing RAN split options do not fully harness data potential, presenting an opportunity to reduce operational expenditures. This paper aims to reduce energy consumption and operational costs in 5G RANs by effectively leveraging network data through two key contributions. First, a Curated Collaborative Learning (CCL) framework that ensures selective collaboration among correlated data sources for accurate traffic and user count forecasting. CCL optimally determines whom, when, and what to collaborate with, significantly outperforming state-of-the-art approaches, including global, federated, personalized federated, and cyclic institutional incremental learnings by 43.9\%, 39.1\%, 40.8\%, and 31.35\%, respectively. Second, the Distributed Unit Pooling Scheme (DUPS) is proposed, leveraging deep reinforcement learning and prediction inferences from CCL to reduce the number of active DU servers efficiently. DUPS dynamically redirects traffic from underutilized DU servers to optimize resource utilization. This improves energy efficiency by up to 89\% over conventional strategies, where all DU servers remain turned on regardless of load, thereby translating into substantial monetary benefits for operators. By integrating CCL-driven predictions with DUPS, this paper demonstrates a transformative approach for minimizing energy consumption and operational costs in next-generation mobile networks, significantly enhancing efficiency and cost-effectiveness.

\end{abstract}

\begin{IEEEkeywords}
Edge architecture, distributed edge learning, collaborative learning, network traffic prediction, DU pooling, sustainable RAN.
\end{IEEEkeywords}

\section{Introduction}

\IEEEPARstart{M}{ulti-Access} Edge Computing (MEC) is a key enabler of Ultra-Reliable Low Latency Communication (URLLC) services in Beyond 5G (B5G) networks, by placing computing capabilities at the network edge \cite{r1, r2,r3,r4,r5}. In modern Radio Access Networks (RAN), MEC integrates with the network architecture to bring computation closer to users, enabling fast and efficient processing. Supporting this, the Open Radio Access Network (ORAN) specification established a split of the traditional Baseband Unit (BBU) into a Central Unit (CU) and Distributed Unit (DU) \cite{r6}. The CU, located in Edge Clouds (ECs), manages non-real-time functions such as packet routing, while the DU, positioned at each Base Station (BS) alongside the Radio Unit (RU), manages real-time tasks such as signal processing, ensuring URLLC services. However, keeping DU servers active at all BS during low-traffic periods leads to unnecessary energy consumption and, consequently, higher operating expenses (OpEx) for operators. This creates an opportunity to reduce OpEx by redirecting traffic from low-traffic DU servers to the nearest available servers, capable of handling the traffic load while maintaining latency requirements. This requires an efficient traffic prediction framework that integrates with the distributed B5G architecture and outperforms existing learning approaches.

\begin{figure}[t]
\centering
\includegraphics[width=3.5in]{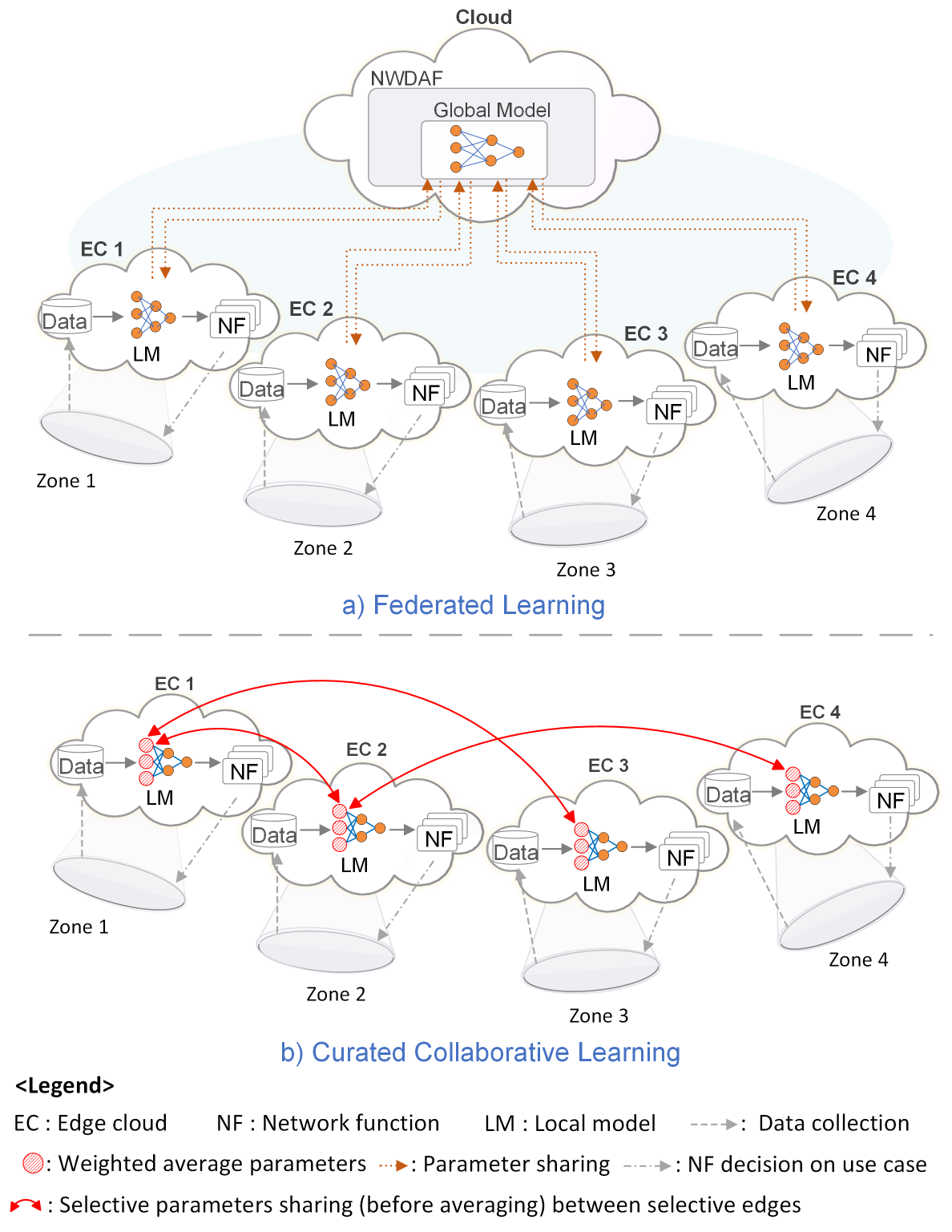}
\caption{ Conceptual framework comparison of FL and CCL.}
\label{fig_1}
\end{figure}

A common approach to traffic prediction relies on a centralized Global Learning (GL) that aggregates data from all edge nodes. While straightforward, it suffers from limitations like scalability and privacy due to high data transmission overhead \cite{r3}. Individual Edge Learning (IdEL) mitigates this by training local models at each EC, enabling fast, localized decisions but at the expense of network-wide accuracy. Federated Learning (FL) improves prediction while preserving privacy by collaboratively training a shared model without transferring raw data, as shown in Fig. \ref{fig_1}(a). However, FL struggles with non-IID data in B5G/6G networks, resulting in degraded global model performance \cite{r4, r5, r6, r7}. Cyclic Institutional Incremental Learning (CIIL) trains models sequentially across institutions but with a risk of catastrophic forgetting and failure to capture asymmetric traffic patterns \cite{r9, r10}. These limitations highlight the need for a novel learning approach tailored to the challenges of B5G distributed architecture and traffic prediction.


This paper proposes a unified prediction–control pipeline for energy-efficient RAN operation, in which traffic prediction is performed using Curated Collaborative Learning (CCL) and the resulting forecasts are used by DU pooling for proactive control decisions. CCL builds on the Collaborative Learning (CL) framework proposed in \cite{mymass}, which facilitates the sharing of partial trainable model parameters among selected edge models, eliminating the need for a centralized global model for real-time inferences. However, unlike mobility prediction, where collaboration among distant edges improves performance, traffic patterns across edges tend to be similar, making direct collaboration less effective. Instead, here the edges are further divided into logical clusters based on unique traffic pattern variations, and only clusters with high correlation collaborate. To reflect this more refined, cluster-level collaboration, the term ``Curated'' is coined for the first time in this paper. The CCL framework is structured around three key questions: whom, when, and what to collaborate.


\begin{itemize}

    \item ``Whom'' selects collaborators based on traffic correlation, ensuring that each logical cluster collaborates exclusively with the members of its established collaboration set \( C \).

    \item ``When" determines the optimal collaboration frequency using validation loss, as both excessive and insufficient collaboration can degrade the prediction performance of the model.
    
    \item ``What" identifies the model parameters to share with members of the collaboration set for maximum performance. In CCL, only the parameters of the initial model layers are shared, while the remaining layers are trained locally, balancing generalization and local adaptation.

\end{itemize}

CCL has been validated using a real-world cellular traffic dataset from a major operator in Pakistan. It outperforms other methods, including GL, IdEL, FL, Personalized FL (PFL), and CIIL, achieving accuracy improvements of 28.2\%, 32.1\%, 39.2\%, 33.5\%, and 29.7\%, respectively. These results demonstrate its effectiveness for traffic prediction and DU pooling.

Leveraging traffic prediction from CCL, a DU Pooling Scheme (DUPS) is proposed to optimize energy consumption in RAN, where the DU pooling policy directly uses the next interval traffic and user forecasts from CCL, making DUPS functionally dependent on these outputs. In DUPS, a deep reinforcement learning (DRL) agent at each BS decides whether to switch a DU server on or off. A centralized critic evaluates the overall reward based on factors such as energy consumption, latency, and penalties for violations. The critic then proportionally distributes the advantages among the agents according to their contributions to the total reward. This adaptive approach dynamically adjusts the number of active DU servers according to the traffic demand at each network edge, significantly reducing energy consumption during low-traffic periods. Additionally, the DUPS also employs a routing mechanism to efficiently redirect traffic from deactivated DU servers to the nearest active ones, ensuring smooth resource utilization. Using the same dataset, the DUPS is benchmarked against conventional strategies. The results demonstrate that the DUPS effectively reduces the number of active DU servers, consequently reducing energy consumption by up to 89\%, and ultimately, reducing OpEx.

The major contributions of this paper are summarized below.

\begin{itemize}

     \item An AI-driven unified prediction-control pipeline is proposed for a distributed edge architecture, where traffic and user count are predicted by CCL, and DUPS uses these forecasts for resource availability-aware DU pooling to reduce RAN energy consumption while satisfying latency constraints. Energy consumption has been reduced by up to 89\% during low-traffic hours compared to the state-of-the-art approaches.

    \item A strategic CCL mechanism using multitask DL models in Edge Clouds ECs, i) exploiting similarity in traffic patterns to determine whom to collaborate with, ii) observing training progress to set when to collaborate, and iii) defining what to collaborate. The novelty improves the prediction performances by an average of 33\% compared to other learning approaches.

    \item A multi-agent DRL framework at each EC that, using the predictions from the CCL, establishes a trade-off between energy consumption, latency, and rejection rate by selecting optimal DU pools.

    \item A comprehensive performance evaluation is conducted using a real-world network traffic dataset from a large city-wide mobile network. Results demonstrate that the proposed CCL and DUPS achieve lower Mean Absolute Error (MAE) and reduced energy consumption, respectively, compared to their respective state-of-the-art counterparts. 
\end{itemize}

The paper is organized as follows: Section 2 provides a brief overview of existing learning frameworks, related studies on traffic prediction, and academia and industry literature on DU pooling. Section 3 details the proposed CCL for traffic forecasting and its integration into the DUPS. Section 4 evaluates the performance of CCL for traffic prediction and DUPS, comparing them with their respective state-of-the-art. Finally, Section 5 concludes the paper.

\section{Related studies}
This section provides a comprehensive overview of various learning approaches widely proposed for B5G networks, followed by recent state-of-the-art studies targeting traffic forecasting using any of the mentioned learning approaches. Finally, the related work for DU pooling strategies by academia and industry is discussed.

\subsection{Distributed Learning Approaches}
Machine Learning (ML) traditionally relied on centralized ``big data'' training, where data was collected and processed in the cloud. Although straightforward, this approach incurs high communication costs and privacy risks. To overcome these limitations, distributed ``small data'' processing at the edge has gained traction, enabling real-time ML training with minimal data transfer. FL facilitates this by allowing agents to collaboratively train models while keeping data local, preserving privacy, and maintaining accuracy \cite{r20, r200}. Multi-agent reinforcement learning builds on FL by enabling distributed decision-making through deep networks \cite{r21}, but its application in large-scale wireless systems is hampered by significant communication overhead. Personalized Federated Learning (PFL) addresses the limitation of non-IID data by enabling federated multitask learning \cite{r22, r89}, allowing devices to perform correlated but distinct tasks. Despite these advances, there remains a need for adaptive and efficient distributed learning methods that maximize data utility while ensuring privacy.


To further improve the efficiency of distributed edge environments, the authors of \cite{r9} investigated data-private CL methods in the medical science domain, adopting a sequential training approach across institutions rather than parallel training. In this method, each institution trains the model before passing it to the next, continuing until all have completed the process. However, this serial approach can lead to catastrophic forgetting, where the model favors the most recently seen data \cite{r10}. An effective alternative involves iterative computation of model weight updates from batches that combine data across multiple non-IID institutional datasets, rather than averaging updates from institution-specific batches \cite{r23}. A recent paper proposes a collaborative distributed Network Data Analytics
Function (NWDAF) architecture for B5G, enhancing NF data localization, security, and training efficiency while reducing control overhead and improving accuracy through real-time network data testing \cite{r24}. However, this architecture includes additional global model training logical functions for each NF, which increases the network overhead and cost. In particular, to the best of our knowledge, no prior work has addressed network traffic prediction within the CL architecture.

\subsection{Traffic Forecasting}

ML techniques have gained significant attention for network traffic prediction, leveraging models such as ensemble methods and tree-based algorithms. For instance, an ensemble strategy incorporating bagging and a light gradient boosting machine has been proposed to enhance the prediction accuracy \cite{r25}. Similarly, the sensitivity of ML-based traffic prediction models to dataset variations has been highlighted, emphasizing the impact of data quality on model performance \cite{r26}. Furthermore, linear ML models and ensembles have been explored for mobile traffic prediction, demonstrating the need for data preprocessing despite their interpretability and efficiency \cite{r27}. However, traditional ML models struggle with complex traffic patterns and datasets, which require more advanced techniques.


To overcome these limitations, researchers have adopted deep learning, particularly time-series models such as Recurrent Neural Networks, Long Short-Term Memory (LSTMs), and Gated Recurrent Units (GRUs), to capture temporal dependencies in network traffic \cite{r28, r29}. An enhanced GRU model has also been used for satellite network traffic prediction \cite{r30}. However, since traffic data involves spatial correlations, time-series models alone are insufficient. Convolutional approaches, such as graph convolution with LSTM, improve accuracy by capturing spatiotemporal patterns \cite{r31}. Likewise, CNN-LSTM and CNN-GRU models have proved effective in environments with multiple access points \cite{r32, r33}. These studies demonstrate that while convolutional deep learning methods improve prediction accuracy, optimizing training strategies for real-world data remains essential. Ultimately, accurate traffic forecasting is crucial for operators to optimize DU server allocation and reduce OpEx efficiently.

\subsection{DU Pooling}

The disaggregation of traditional BBUs into separate RU, DU, and CU components in ORAN is a fundamental shift in 5G networks, enabling flexible configurations tailored to service requirements \cite{r2}. The RU manages radio frequency functions, while the DU, deployed on Commercial Off-The-Shelf (COTS) servers at BS sites, executes real-time Layer 1 and lower Layer 2 functions, ensuring low latency. The CU, responsible for non-real-time higher Layer 2 and Layer 3 functions, centralizes control over multiple DUs. Communication between these components occurs via the FrontHaul (FH), MidHaul (MH), and BackHaul (BH) links, facilitating functional splits that enhance centralization, virtualization, and optimized placement decisions \cite{r34}.

Researchers have investigated DU and CU placement strategies, often neglecting operational costs and functional split constraints. Traffic-aware CU pooling has been proposed to reduce power consumption while keeping DUs at each BS, employing Mixed Integer Linear Programming (MILP) and heuristic algorithms with limited improvements \cite{r35, r35a}. An optimized routing and packet scheduling framework within an O-DU pool enhances processing efficiency for URLLC and enhanced Mobile Broadband (eMBB) services \cite{r36}, while reinforcement learning improves DU-CU placement in optical transport networks, outperforming heuristic methods in resource allocation \cite{r37}. Samsung highlights that decoupling DU functions from hardware mitigates load imbalances, reducing excess resource allocation \cite{r38}, whereas Fujitsu advocates dynamic resource scaling based on traffic variations, significantly reducing power consumption during low-traffic periods \cite{r39}. These insights align with our proposed system, emphasizing efficient DU pooling, adaptive resource management, and cost-effective RAN deployment.


\begin{figure*}[t]
\centering
\includegraphics[width=7in]{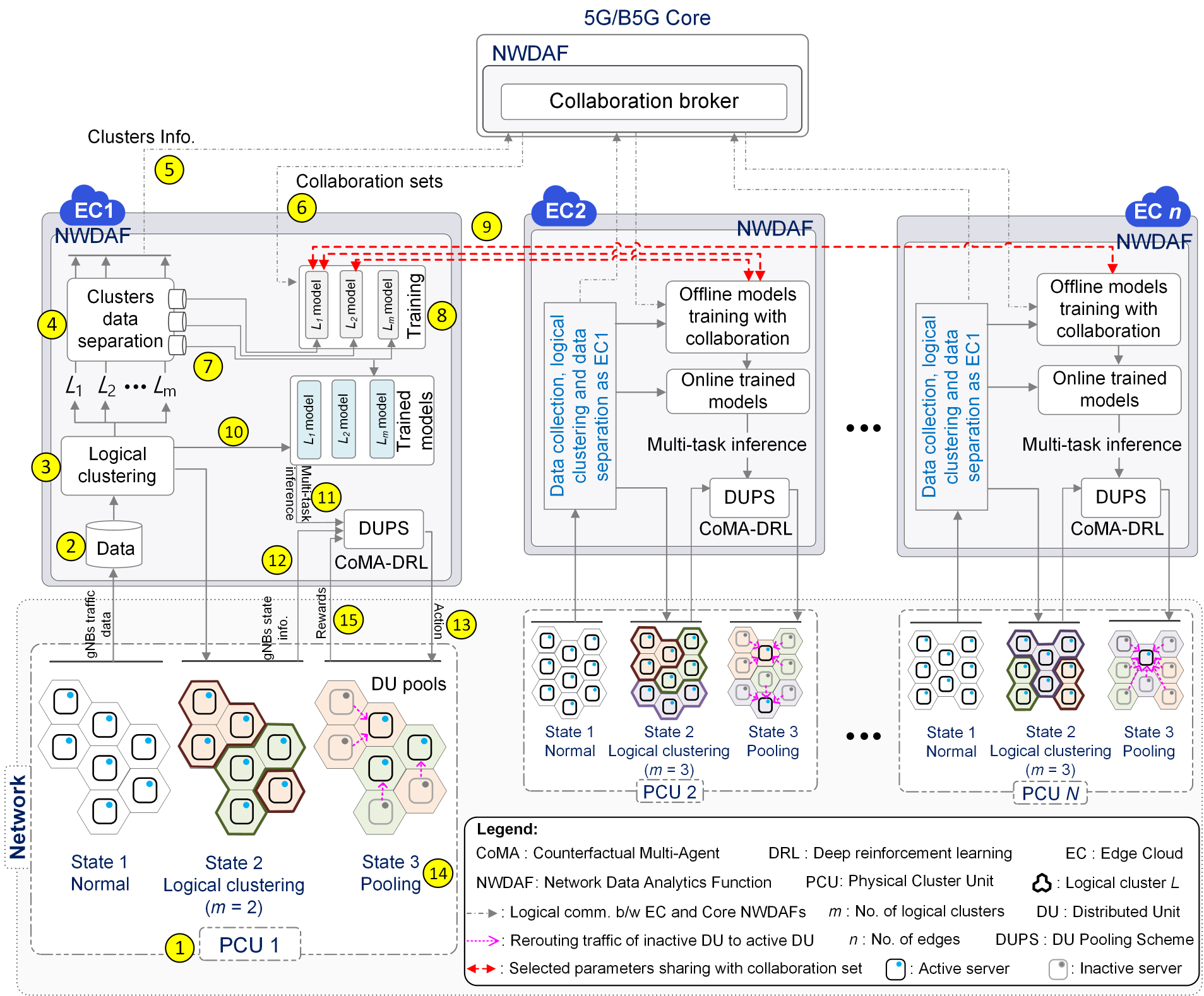}
\caption{Curated Collaborative Learning and DRL-based DU pooling architecture in 5G/B5G.}
\label{fig_2}
\end{figure*}

\section{Curated Collaborative Learning based Traffic Forecasting and DRL-based DU Pooling}

This study proposes a unified prediction--control pipeline for 5G/6G edge architectures, in which traffic and user forecasts drive proactive DU pooling decisions. Within this pipeline, the CCL framework incorporates multitask learning to simultaneously predict traffic and user count at BSs. The DRL-based DUPS module uses these forecasts to dynamically reduce the number of active DU servers with low or no traffic, significantly lowering OpEx. 
Furthermore, a traffic redirection mechanism is included to route traffic from deactivated DU servers to the nearest active ones with sufficient capacity to handle the load.

\subsection{System Architecture}
This section provides the overall B5G system architecture, illustrated in Fig. 2, which includes traffic prediction and DU pooling operations as sequential stages of a unified prediction–control pipeline. In this pipeline, the DU pooling decisions at each EC are driven by the next-interval traffic and user forecasts generated by CCL. The NWDAF collects digital traffic records and categorizes the BSs into Physical Cluster Units (PCUs) \circled{1} based on their geographic proximity. PCUs forward data to their respective ECs for processing \circled{2}. Each data record is represented as:  

\begin{equation}
\setcounter{equation}{1}
d_{t,j,i} = \langle t,\ j,\ i,\ \text{latitude}_i,\ \text{longitude}_i,\ u_{i,t},\ \tau_{i,t} \rangle
\label{eq:data_record}
\end{equation}

\noindent where $t$ is the timestamp, $j$ is the edge ID, $i$ is the BS ID, $(\text{latitude}_i,\ \text{longitude}_i)$ is its location, $u_{i,t}$ is the number of users, and $\tau_{i,t}$ is the traffic volume in GB. The complete dataset is defined as:  


\begin{equation}
\mathcal{D} = \left\{ d_{t,j,i} \mid j = 1,\dots,k;\ i = 1,\dots,n_j;\ t = 1,\dots,T \right\},
\label{eq:dataset}
\end{equation}

\noindent where $\sum_{j=1}^{k} n_j = n$, $k$ is the number of edges, $n_j$ is the number of BSs in edge $j$, $n$ is the total number of BSs across all $k$ edges, and $T$ is the number of time intervals. Using Urbanflow Peak-time Clustering (UPC), which first groups BSs by peak-hour traffic patterns and then consolidates them based on traffic similarity, the NWDAF groups BSs into logical clusters and assigns each cluster a unique logical ID \circled{3}. The clustered data, i.e., time-series tuples from multiple BSs, are divided into logical clusters \circled{4}, and also sent to the 5G core network \circled{5} to form collaboration sets $C$ for collaboration during model training. The core network calculates the correlations among the clusters across the ECs and sends the collaboration sets back to the ECs \circled{6}.

    Edge computing, including model training and inference, is performed at the EC that hosts the distributed NWDAF, AMF, and other 5G functions. Neither the user devices nor the BSs carry out these tasks. User devices act only as data generators, and BSs serve as access points that forward traffic to the EC, where computation-intensive processes such as training, inference, and DU server control are executed. Each logical cluster then uses its data \circled{7} to train models \circled{8} and collaborates with other clusters in its $C$ during the training process \circled{9}. After offline training, the trained models are used for real-time inference \circled{10}. The inference \circled{11}, and the environment state \circled{12} are fed into the DRL agent for the DUPS, which recommends an action about which DU servers to activate or deactivate \circled{13}. Traffic from deactivated DUs is rerouted to active ones \circled{14}. The actions of agents are assessed by monitoring energy consumption with latency and capacity constraints, and the rewards are distributed accordingly \circled{15}. This approach optimizes network efficiency by turning off underutilized DU servers, pooling traffic to active servers, and adjusting policies for the next cycle.

The first crucial step in the architecture is logical clustering, which groups BSs based on traffic similarities rather than geographic proximity. The rationale for not using PCUs for collaboration is that location-based clusters often overlook relevant traffic patterns of individual BSs. In contrast, logical clusters ensure that prediction models learn from BSs with truly similar trends, improving the overall prediction accuracy. Therefore, the UPC method is employed to develop logical clusters at each edge. In the first step, each BS $b$ is assigned to its peak-traffic cluster $P_i$. This is done by identifying the hour $i$ when the traffic for a BS is at its maximum, and then grouping all BSs with similar peak hours into the corresponding cluster $P_i$. For instance, three BSs $b_1$, $b_2$, and $b_3$ have their peak traffic at 10 AM, and they are assigned to $P_{10}$. To create a representative profile for each cluster \( P \in \mathcal{P}\), where $\mathcal{P}$ represents the set of all the $P_i$ in an edge, the average traffic is calculated across all BSs in that $P$. For instance, in \( P_{10} \), if \( b_1 \) has traffic values of \([80, 90, 75, \dots]\), \( b_2 \) has \([85, 95, 85, \dots]\), and \( b_3 \) has \([90, 100, 80, \dots]\), then the average traffic is \([85, 95, 80, \dots]\), forming a representative profile for calculating correlation. The pseudocode for this process is provided in Algorithm S1 in the supplementary materials.

In the next step, the Pearson Correlation Coefficient (PCC) value between all peak-traffic-based clusters $P_i$ is calculated to quantify the similarity in traffic patterns among all $P \in \mathcal{P}$. The PCC measures the linear correlation between two variables, in this case, the traffic patterns of $P$. The value of this coefficient ranges from -1 to +1, where -1 indicates a high negative correlation, 0 indicates no correlation, and +1 indicates a high positive correlation between the $P_i$ and $P_j$. The PCC value $p_{x,y}$ for any two variables $x$ and $y$ is calculated using Eq. (\ref{eq1}).

\begin{equation}
p_{x,y} = \frac{{}\sum_{i=1}^{h} (x_i - \overline{x})(y_i - \overline{y})}
{\sqrt{\sum_{i=1}^{h} (x_i - \overline{x})^2  \sum_{i=1}^{h}(y_i - \overline{y})^2}}
\label{eq1}
\end{equation}

\noindent where, $\overline{x}$ and $\overline{y}$ represent the arithmetic mean of the hourly traffic for $P_{i}$ and $P{j}$ over $h$ hours, respectively. Afterward, all \( P \) clusters are consolidated into \( m \) logical clusters within each edge, aiming to maximize the mean PCC within each logical cluster \( L_i \). For instance, if \( |\mathcal{P}| = 3 \), say \( P_1 \), \( P_2 \), and \( P_3 \), there are multiple ways to form \( m \) logical clusters. Suppose \( m = 2 \), meaning two logical clusters, \( L_1 \) and \( L_2 \). Using these  configurations where \( L_1 = \{P_1\} \) and \( L_2 = \{P_2, P_3\} \). In this case, the self-correlation within \( L_1 \) is 1, and the pairwise correlation within \( L_2 \) is 0.8, yielding an average PCC of 0.9. In another configuration, let \( L_1 = \{P_1, P_2\} \) and \( L_2 = \{P_3\} \). Here, the pairwise correlation within \( L_1 \) is 0.5, and the self-correlation within \( L_2 \) is 1, resulting in an average PCC of 0.75, which is lower than the previous configuration. Similarly, the combination \( L_1 = \{P_1, P_3\} \) and \( L_2 = \{P_2\} \) yields a lower PCC value of 0.6. Thus, the final logical clustering will use the configuration that yields the highest average PCC, which in this example is the first combination of \( L_1 = \{P_1\} \) and \( L_2 = \{P_2, P_3\} \). 


\subsection{Curated Collaborative Learning}
The proposed collaboration strategy in the architecture is structured into three steps, with details outlined in Algorithms \ref{algo1} and \ref{algo2}.

\begin{algorithm}[t]
\caption{Construction of collaboration set}
\small
\LinesNumbered
\SetKwInOut{Input}{Input} 
\SetKwInOut{Output}{Output}
\label{algo1}

\Input{List of logical clusters $\mathcal{L}$ of every edge; hours $H = \{0, \dots, 23\}$; correlation threshold for logical cluster $\psi$.}
\Output{A collaboration set for each logical cluster.} 
\BlankLine

$\mathbb{L}=\bigcup_{\forall\mathcal{L}}\mathcal{L}$ \\
\tcp{Whom to collaborate with?}

\For{each logical cluster $L_i \in \mathbb{L}$}{
    \For{$h \gets 0$ to $|H|-1$}{
        $L_i.\text{traffic}(h) \gets \frac{\sum_{\forall b \in L_i}{b.\text{traffic}(h)}}{|L_i|}$ \\
    }
}
\BlankLine
\tcp{Calculating pairwise PCC among all $L_i$}

\For{$i \gets 0$ to $|\mathbb{L}|-1$}{
    \For{$j \gets 0$ to $|\mathbb{L}|-1$}{
        $r_{i,j} \gets$ PCC between $L_i$ and $L_j$ using Eq. (\ref{eq1}) \\        
    }
}
\BlankLine
\For{$i \gets 0$ to $|\mathbb{L}|-1$}{
   $C_{i} \leftarrow \emptyset$ \tcp{Initialize collaboration set}
    \For{$j \gets 0$ to $|\mathbb{L}|-1$}{
        \If{$i\neq j$ and $r_{i,j} \geq \psi$}{
            $C_{i} \leftarrow C_{i} \cup \{ L_j\}$
        }
    }
}
\Return $\{C_{i} \mid i \in \mathbb{L}\}$ \tcp{Return the collaboration set for each logical cluster.}
\end{algorithm}

\begin{algorithm}[!t]
{
\caption{Collaborating frequency and parameters}
\small
\LinesNumbered
\SetKwInOut{Input}{Input} 
\SetKwInOut{Output}{Output}
\label{algo2}

\Input{Collaboration set $C$ for logical cluster $L$; collaboration frequency threshold $\omega$; validation loss $v$; maximum epochs $T$.}
\Output{Updated parameters $W$.} 
\BlankLine

\While{1}{ 
 \For {each epoch $i$}{
    $v \gets \frac{1}{2 \times |batch|} \sum_{data \in batch} \Big(\text{MAE(pred.traffic, } data.\text{traffic)} + \text{MAE(pred.users, } data.\text{users)} \Big)$
    
    \If{$|\nabla v| \leq \omega$ or $i > \frac{T}{2}$}{
        \textbf{break} \tcp{Stop current training and wait for collaboration.}
    }
 }
\BlankLine
\tcp{Collaborating parameters}
\For {each logical cluster $L_i \in C$}{
    $\alpha_i \gets \frac{|L_i|}{|L|}$
}
$W \gets \frac{W+\sum_{L_i \in C}{\alpha_i \times W_i}}{|C|+1}$
}
\Return $W$ \tcp{Return the updated parameters.}
}
\end{algorithm}

\subsubsection{Whom to Collaborate With? (Setting $C$)} 
The essence of this study lies in its collaborative approach, beginning with the critical task of selecting collaborators and forming a collaboration set $C$ at the core network, as stated in Algorithm \ref{algo1}. When predicting the BS traffic, there is a considerable likelihood that two edges exhibit similar traffic patterns, which do not accurately reflect the actual differences between the traffic of individual BSs. Therefore, it is more effective to collaborate at the level of logical clusters rather than at the level of entire ECs. The core network computes the pairwise PCC among all $L \in \mathcal{L}$, as in lines 1-7. 
For each logical cluster $L_i$, if the value of $r_{i,j}$ with any $L_j \in \mathcal{L}$ exceeds a given threshold $\psi$, then $L_j$ is added to the collaboration set ${C_i}$ for $L_i$, as in lines 8-12. This selective collaboration ensures that the model learns exclusively from correlated models of logical clusters across all edges. Such selective refinement enhances the resilience and adaptability of the model by accommodating out-of-distribution patterns found in any edges from $C$ but not present in the source model. It is worth noting that this framework can also be applied to the level of EC collaboration in other prediction tasks when there is a dissimilarity between the task patterns of the edges \cite{mymass}.




    


\subsubsection{When to Collaborate? (Setting $\omega$)} 
Sharing parameters before the models have learned meaningful patterns leads to unnecessary overhead and computational inefficiencies. Therefore, determining the right collaboration frequency threshold $\omega$ is crucial. It is proposed that each collaborating model shares trainable parameters only after learning certain patterns. This could be monitored through $\nabla v_{i}$, the rate of change in the validation loss of the individual logical cluster model $L_i$ while they train on their data (lines 1-7 of Algorithm 3). For instance, if $\nabla v_{i}$ fell below a given threshold $\omega$, the model is deemed ready to share its trainable parameters. This approach reduces overhead while maintaining a satisfactory performance.

\subsubsection{What to Collaborate?} 
    During offline training, each model trains on its local dataset while selectively sharing the parameters of the initial layers of the logical cluster with collaborators. Extensive evaluations, including ablation studies in \cite{mymass}, demonstrated that collaborating on the initial layers significantly improved generalization. This is because the initial layers are less affected by backpropagation and thus better suited for collaboration. After receiving parameters from collaborators, the source NWDAF model takes a weighted average of the shared parameters and updates the corresponding layers (Algorithm 3, lines 8-10). This approach allows the initial layers to learn from broader patterns across clusters, while the deeper layers continue to learn from local data. It is important to note that the parameters are shared before averaging, ensuring that collaborators receive only the direct contribution from the source model, which helps mitigate the influence of models outside their collaboration set and leads to more effective learning.

\begin{figure*}[t]
\centering
\includegraphics[width=6in]{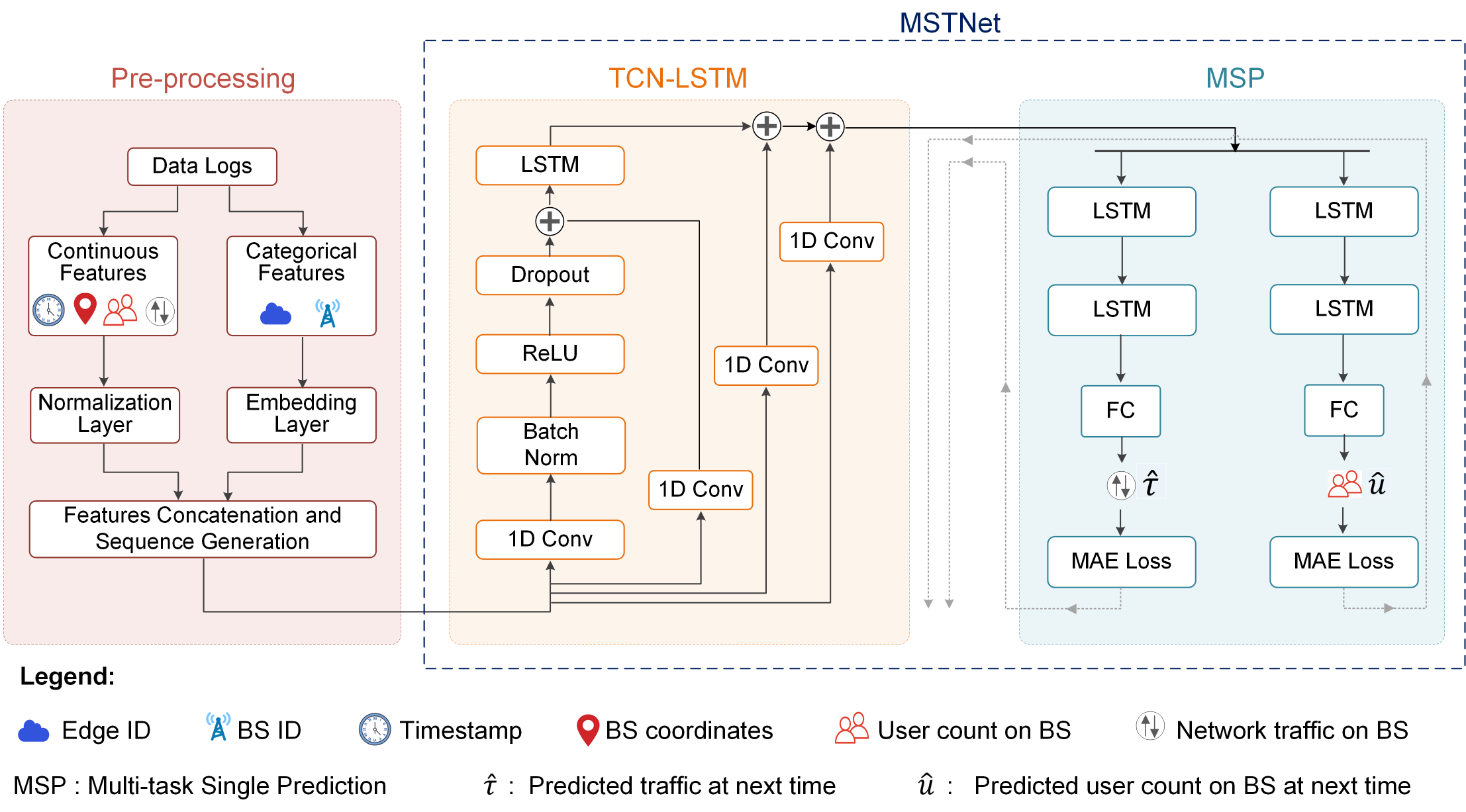}
\caption{Multitask Single-step Temporal Network (MSTNet) model for predicting \redst{BS} the next traffic $\hat{\tau}$ and user count $\hat{u}$ at the BS.}
\label{fig_2a}
\end{figure*}

While resembling split learning, the key difference lies in the active involvement of all logical clusters in parameter sharing in split learning. Therein, the local model of the logical cluster $L_x$, for instance, receives contributions from all models, potentially compromising its ability to capture local data nuances and performance. On the other hand, the proposed approach shares parameters intelligently between selected clusters within the ECs, which helps them to learn better collectively. This tailored approach improves how accurately BS traffic can be predicted, which can help develop efficient resource allocation strategies.

\subsection{Multitask Single-step Temporal Network for CCL}
MSTNet, illustrated in Fig. \ref{fig_2a}, is designed for simultaneous traffic and user count prediction at BSs that encompasses three key components: TCN-LSTM, Residual Connections (RC), and Multitask Single-step Prediction (MSP). TCN-LSTM extracts complex temporal patterns from mobile radio and location data, while RC ensures the retention of long-term dependencies. MSP, structured with two branches, enables joint learning of BS traffic $\hat{\tau}$ and user count $\hat{u}$, enhancing predictive performance by leveraging their interdependencies.

The TCN layer employs 1D dilated causal convolutions to efficiently capture short-term dependencies without increasing the model complexity. Batch normalization stabilizes training, while Rectified Linear Unit (ReLU) activation enhances nonlinearity, improving pattern extraction. A 20\% dropout rate is applied before the LSTM layer that captures long-range dependencies to prevent overfitting. Unlike traditional activations that suffer from gradient saturation, ReLU ensures stable backpropagation, making the model robust in learning both short- and long-term correlations. The RC further enhances the MSTNet feature extraction by linking intermediate and final outputs within the TCN-LSTM blocks. These connections allow high-dimensional features to be processed early while refining lower-dimensional representations later, preventing information loss and gradient vanishing. This structured information flow improves the ability of the model to capture complex patterns.

The MSP block finalizes the architecture with two LSTM-based branches, each followed by Fully Connected (FC) layers for traffic and user count prediction. The traffic branch consists of LSTM layers with 256 and 32 units, followed by a hidden layer with 16 nodes and a single output node. The user count branch follows a similar structure but includes a hidden layer with 8 nodes. Both branches optimize MAE loss functions, given by Eq. (\ref{eq2}), ensuring precise predictions. Sequential training integrates the losses from both branches, refining the TCN-LSTM block for optimal performance across both tasks.

\begin{equation}
    \ell_{\omega} = \frac{1}{N} \sum_{i=1}^{N} |\omega_{q+1} - \hat{\omega}_{q+1}|    
    \label{eq2}
\end{equation}
\noindent
where, $\omega_{q+1}$ represents the ground truth, while $\hat{\omega}_{q+1} $ is the prediction for the next hour (${q+1}$). $\omega$ is either $\tau$ or $u$ to predict BS traffic or user count, respectively. Moreover, the selection of the multitask design over two separate single-task models is further supported by experimental evidence, as reported in the ablation studies in the supplementary material.

\subsection{DRL for DU pooling}
Leveraging the most accurate traffic prediction, the proposed DUPS employs a Multi-Agent DRL (MADRL) framework to optimize energy efficiency and minimize service delay in RANs. The MADRL is particularly well-suited for the DUPS due to its ability to handle multiple agents (DUs) in a decentralized manner, enabling dynamic resource management and adaptability to fluctuating traffic conditions. The problem is formulated as a Partially Observable Markov Game (POMG), where each DU agent interacts with a partially observable environment to make decisions based on local observations and shared information.

\subsubsection{Problem Formulation}

    The POMG is defined by the tuple $(\mathcal{N}, \mathcal{S}, \mathcal{A}, \mathcal{O}, \mathcal{T}, \mathcal{R}, \gamma)$. Here, $\mathcal{N} = \{1, 2, \dots, n\}$ denotes the index set of agents, where each agent corresponds to one BS. Thus, the number of BSs is given by the cardinality of this set, $|\mathcal{N}| = N$. Each agent controls a DU server and operates independently. The finite state space is denoted by $\mathcal{S}$, where each state $s_t \in \mathcal{S}$ includes the current timestamp $t$, DU server power status (on/off), predicted traffic $\hat{\tau}$, and predicted user count $\hat{u}$ at the BS. The joint action space for all agents is given by $\mathcal{A} = \prod_{i=1}^n a^i$, where $a^i \in \{0,1\}$ indicates whether agent $i$ turns its DU off (0) or on (1). The joint observation space is defined as $\mathcal{O} = \{o_1 \times o_2 \times \dots \times o_n\}$, where each observation $o_i$ includes the current traffic volume and user count at $BS_i$, together with any locally available estimates needed for decision-making. The state transition probability is represented by $\mathcal{T}: \mathcal{O} \times \mathcal{A} \to \mathbb{R}$, which models how the environment transitions based on joint actions, such as rerouting traffic due to DU shutdowns. The reward function $\mathcal{R}: \mathcal{O} \times \mathcal{A} \to \mathbb{R}$ provides feedback to the agents, incorporating energy savings, FH latency, and penalties for service degradation. Finally, $\gamma \in [0, 1]$ denotes the discount factor, where $\gamma = 0$ makes the agent focus only on immediate rewards, while $\gamma$ close to 1 makes it prioritize long-term cumulative rewards.

\subsubsection{Key Components}
The environment consists of a RAN where each BS is managed by a DU controlled by an actor agent. These agents collaboratively address the DU pooling problem by dynamically switching DUs on or off to balance energy efficiency and service performance. At time $t$, the state $s_t = [t, DU_s, \hat{\tau}, \hat{u}]$ includes the current time $t$, the DU power state ($DU_s$), the predicted traffic load ($\hat{\tau}$), and the predicted user count ($\hat{u}$) on the BS from MSTNet. Each agent selects an action $a_t \in \{0, 1\}$, where $a_t = 0$ turns the DU off and $a_t = 1$ turns it on. This action influences both the next state and the reward received. The reward function $r_t$, given by Eq. \ref{eq3}, is designed to balance energy efficiency, service delay, and penalties.
\begin{equation}
r_t = \alpha E_t - (1-\alpha) D_t + \beta P_t
\label{eq3}
\end{equation}

\noindent
Here, $E_t$ represents energy consumption, which is 70\% at no load and scales linearly to 100\% at full load. $D_t \in [0, 1]$ is the normalized FH latency: it is 0 when the BS processes traffic locally, and increases linearly as redirection latency increases, reaching 1 at the 75\,$\mu$s threshold, following the upper bound defined for 5G RU–DU split timing [34, 35]. $P_t \in \{0,1,2\}$ is a discrete penalty term: the penalty is triggered if a service constraint is violated. Specifically, $P_t = 1$ is triggered if either (i) the FH latency limit is breached or (ii) the DU server receiving redirected traffic is overloaded (i.e., its total traffic exceeds its capacity). $P_t = 2$ is triggered if both violations occur in the same timestep; otherwise, $P_t = 0$. Such penalty functions are widely used in constrained DRL for 5G resource management [36,37]. This design allows $D_t$ to provide smooth, continuous feedback for latency minimization, while $P_t$ enforces strict penalties for hard constraint violations.  
The weights $\alpha$, $(1 - \alpha)$, and $\beta$ enable the agent to trade off energy efficiency, latency, and constraint satisfaction during training.



\begin{algorithm}[t]
{

\caption{COMA-based DU pooling}
\small
\SetKwInOut{Input}{Input}
\SetKwInOut{Output}{Output}
\label{algo3}

\Input{Network insights including timesteps $t$; BS IDs; predicted traffic $\hat{\tau}$; predicted user count $\hat{u}$; $n$ BSs, each with one active DU server; maximum steps per episode $T$; number of episodes $e$; buffer memory $M_B$ with capacity $B$.}
\Output{Trained policy network $\pi_\theta$.}
\BlankLine

Initialize $n$ DUs for $n$ BSs, critic network $\theta_{\text{critic}}^1$, target critic network $\hat{\theta}_{\text{critic}}^1$, actor network $\theta_{\text{actor}}^1$, and $M_B$.\\
\tcp{\(\lambda\): Bias-variance tradeoff parameter in Q-value estimation.} 
\tcp{\(\alpha\): Learning rate for updating network parameters.} 
\tcp{\(\gamma\): Discount factor for future rewards.}

\For{each episode $i \gets 1$ to $e$}{ 

    \tcp{Step 1: Trajectory collection}
    \For {each timestep $t \gets 1$ to $T$} {
        Each actor with $\theta_{\text{actor}}$ selects DU status action $a_t^z \in \{0,1\}$ for DU $z$, based on state $S_t$. \\ 
        \If {$a_t^z = 0$}{ 
            Execute traffic redirection via Algorithm 5. \\
        }
        Compute reward $r_t$ using Eq. (\ref{eq3}). \\ 
        $M_B \leftarrow M_B \cup \{t, S_t, A_t, r_t\}$. \\ 
    }

    \tcp{Step 2: Critic network update}
    \For{each timestep $t \gets 1$ to $T$}{
        Unroll states, actions, and rewards from \(M_B\).\\
        Compute target Q-value \(y_t^{(\lambda)}\): \\
        \(y_t^{(\lambda)} = (1 - \lambda) \sum_{n=1}^\infty \lambda^{n-1} G_t^{(n)}\), where \(G_t^{(n)} = r_{t+1} + \gamma r_{t+2} + \dots + \gamma^{n-1} r_{t+n} + \gamma^n V(S_{t+n})\).\\
        \tcp{\(G_t^{(n)}\): \(n\)-step return with \(V(S_{t+n})\) as the value function from the critic network.}
    }
    
    \tcp{Step 3: Target critic synchronization}
    \For{each timestep $t \gets T$ to $1$}{
        Compute gradient for critic network update: \\
        \(\Delta \theta_{\text{critic}} = \nabla_{\theta_{\text{critic}}} (y_t^{(\lambda)} - Q(S_t, A_t))^2\).\\
        Update critic network parameters: \\
        \(\theta_{\text{critic}}^{i+1} = \theta_{\text{critic}}^i + \alpha \Delta \theta_{\text{critic}}\).\\
        Sync target critic network every \(B\) steps: \\
        \(\hat{\theta}_{\text{critic}}^1 \gets \theta_{\text{critic}}^1\).
    }
    
    \tcp{Step 4: Actor network update}
    \For{each timestep $t \gets T$ to $1$}{
        Compute advantage function \(Z_t^j(S_t, A_t)\) for each agent \(j\) using Eq. (\ref{eq4}).\\
        Update actor network gradient: \\
        \(\Delta \theta_{\text{actor}} += \alpha \nabla_{\theta_{\text{actor}}} \log \pi_{\theta_{\text{actor}}}(S_t^j, a_t^j) Z_t^j(S_t, A_t)\).\\
    }
    Update actor network parameters: \\
    \(\theta_{\text{actor}}^{i+1} = \theta_{\text{actor}}^i + \alpha \Delta \theta_{\text{actor}}\).
}

\Return $\pi_\theta$ \tcp{Return the trained policy.}
}
\end{algorithm}

\subsubsection{COMA-based Training}
The Counterfactual Multi-Agent (COMA) algorithm is employed to train the agents. COMA utilizes a centralized critic and distributed actors, enabling efficient policy learning through centralized training and decentralized execution. The centralized critic computes the Q-values, $Q(S_t, A_t)$, for joint state-action pairs and evaluates the advantage function $Z_t^j(S_t, A_t)$ for each agent $j$. The advantage function is defined by Eq. (\ref{eq4}) and it quantifies the contribution of each agent's action to the global reward, enabling the actor network to learn optimal policies.
\begin{equation}
Z_t^j (S_t,A_t ) = Q(S_t,A_t ) - \sum_{a_t^j} \pi^j (a_t^j |S_t^j) \cdot Q(S_t,(A_t^{(-j)},a_t^j)).
\label{eq4}
\end{equation}

Here, it is important to note that at each timestep, all DU agents select their actions simultaneously, forming a joint action vector. The environment then applies the joint DU on/off decisions in a single transition, and performs traffic redirection and latency updates for the DUs that are turned off. A centralized critic evaluates the resulting global state and reward, and credit assignment to individual agents is handled through the COMA advantage function. Therefore, no sequential ordering among agent decisions is required.

\subsubsection{Training Process}
The training process consists of four main steps, as illustrated in Algorithm 3. First, during trajectory collection, actors select actions $a_t^i \in \{0, 1\}$ based on the current state $S_t$. If $a_t^i = 0$, traffic is redirected to the nearest active DU using Algorithm 4. The system transitions to the next state, and the reward $r_t$ is computed using Eq. (\ref{eq3}). Experiences (states, actions, and rewards) are stored in a buffer $M_B$ for subsequent optimization. Second, the critic network is updated by computing target Q-values $y_t^{(\lambda)}$ using the $\lambda$-return method:
\begin{equation}
y_t^{(\lambda)} = (1 - \lambda) \sum_{n=1}^\infty \lambda^{n-1} G_t^{(n)},
\label{eq5}
\end{equation}
where $G_t^{(n)}$ is the n-step return. The critic is updated by minimizing the loss $(y_t^{(\lambda)}-Q(S_t, A_t))^2$.

Third, the actor network is updated using policy gradients and the advantage function $Z_t^j(S_t, A_t)$. The gradient is accumulated over timesteps and applied to optimize the actor network. Finally, the target critic network is periodically synchronized with the main critic network to ensure training stability.

\subsubsection{Network Architecture}
The actor network utilizes the LSTM model to capture temporal dependencies in the state sequence, while the critic network is structured as a feed-forward network with ReLU activation layers and FC layers. Both networks are trained in batch mode with a batch size of 100. The actor and critic networks have learning rates of 0.001 and 0.005, respectively. A target critic network is periodically updated to enhance training stability.

\subsubsection{Traffic Redirection}
Algorithm 4 handles traffic redirection by rerouting traffic from inactive DUs to the nearest active DU, based on the geographical (Euclidean) distance, with sufficient capacity. For each inactive DU $b$, the algorithm identifies the nearest active DU $\hat{b}$ and checks if $\hat{b}$ can accommodate the combined traffic load ($b_j.\hat{\tau} + b_i.\hat{\tau} \leq b_j.\text{capacity}$). If this condition is met, the traffic from $b$ is routed to $\hat{b}$, and $\hat{b}$'s traffic load is updated accordingly.

\begin{algorithm}[t]
\caption{DU pool traffic redirection}
\small
\LinesNumbered
\SetKwInOut{Input}{Input} 
\SetKwInOut{Output}{Output}

\Input{List of all BS $\mathbb{B}$; predicted traffics $\hat{\tau}$ at each DU server $b\in \mathbb{B}$; distance matrix $d(b, b')$ between all DU servers.}
\Output{Routed traffic between DU servers.}

\BlankLine
\For{each base station $b \in \mathbb{B}$}{
    \If{$b.\text{DU} = \text{OFF}$}{
        \While{1}{
            $b' \leftarrow \text{getNextNearestNeighbor}(b)$\;
            \If{$b'.\text{DU} = \text{ON}$}{
                \If{$(b'.\hat{\text{traffic}} + b.\hat{\text{traffic}}) < b'.\text{capacity}$}{
                    \tcp{Route traffic from $b$ to $b'$}
                    $b'.\hat{\text{traffic}} \leftarrow b'.\hat{\text{traffic}} + b.\hat{\text{traffic}}$\;
                    \textbf{break}\;
                }
            }
        }
    }
}
\Return{Updated traffic distribution among DU servers.}
\end{algorithm}

\section{Performance Evaluation}
Since the DU pooling policy relies on the predicted traffic and user count, the performance of DUPS inherently reflects the accuracy of the upstream CCL predictions. Thus, this section evaluates the performance of the two key components of this work. First, for traffic prediction, CCL is compared with several schemes, including GL, IdEL, FL, PFL, and CIIL. Additionally, detailed ablation studies, which explore different variations of CCL along with the validation loss convergence, are provided in the supplementary materials. Moreover, the effectiveness of a single multitask MSTNet over two separate single-task models is also provided as an ablation study in the supplementary materials. Leveraging the predicted traffic and user count from the CCL, DRL-based DUPS is compared with the conventional DU pooling method and MILP scheme for ECs with contrasting traffic conditions. This comparative analysis highlights the improvements by DUPS to improve network sustainability by improving energy efficiency and reducing OpEx. Lastly, the optimal coefficient parameter values used for each EC are determined through experiments detailed in the ablation studies.

\begin{figure}[!t]
\centering
\includegraphics[width=3.5in]{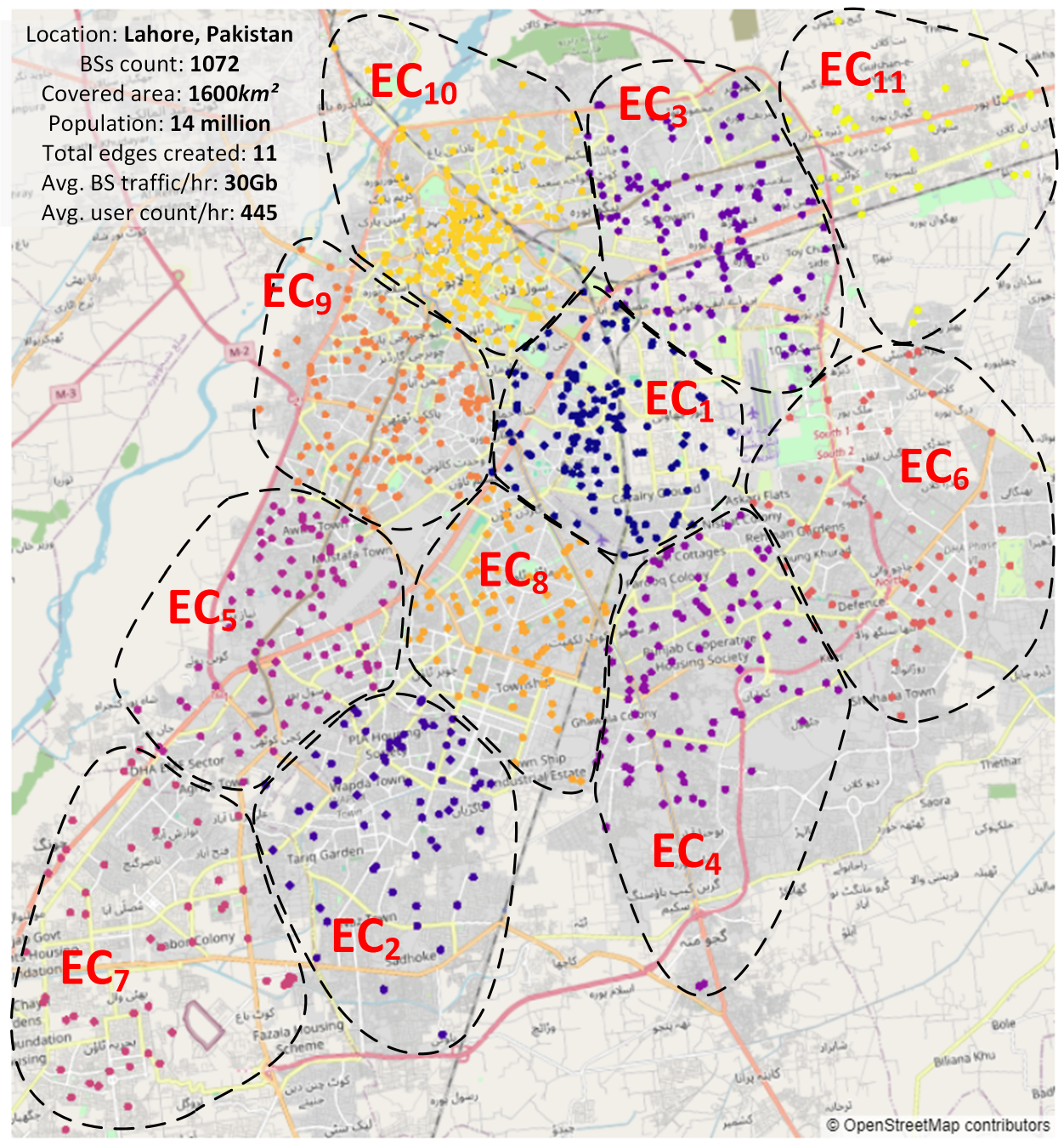}
\caption{BSs assignment to 11 Edge Clouds ECs.}
\label{fig_5}

\end{figure}

\subsection{Dataset and Experimental Setup}
The CCL and DUPS frameworks are evaluated using a proprietary mobile network traffic dataset from a 4G LTE operator in Lahore, Pakistan. The dataset spans 1072 BSs across 1,600 $km^2$, with hourly traffic logs collected over seven days. Based on insights from prior studies \cite{multi, iCHO}, incorporating multivariate features effectively captures diverse traffic variations, ensuring a comprehensive representation of network dynamics over the collected seven-day period. Each record includes the BS ID, sector ID, location, traffic volume, and user count. The missing sector data is corrected via intra-BS averaging, while missing traffic or user values are interpolated using adjacent timestamps. The dataset is scaled using Eq. (\ref{eq:du_scaling}) to adapt the LTE dataset for 5G analysis, taking a 5G BS capacity of 4Gbps \cite{r44, r48}.

\begin{equation}
DU_i = \frac{4\,\mathrm{Gbps}}{\max\!\left\{ \text{UserTraffic}_{DU_i} \right\}}, 
\quad \forall\, DU_i \in \text{Dataset}.
\label{eq:du_scaling}
\end{equation}

Moreover, this cellular network does not employ any ECs in which logical clustering could be performed. Thus, the $k$-means clustering algorithm is utilized first, aiming to allocate $n$ BSs to $k$ ECs. This process is thoroughly explained in the supplementary material and resulted in $k$ = 11 ECs, encompassing 1072 BSs distributed across the city, as shown in Fig. \ref{fig_5}.




    The primary task is to predict BS traffic. However, these 11 ECs cannot be directly used for collaboration due to their nearly identical average traffic patterns, making such collaboration ineffective. A more effective approach is to further subdivide the ECs into logical clusters $L$ based on finer-grained temporal traffic variations. Therefore, each edge is first divided into peak-time clusters according to peak traffic times using UPC. In this process, each BS is assigned to a peak-time cluster based on the hour of the day when it experiences maximum traffic. For example, if a BS reaches its traffic peak at 10 AM, it is grouped into Peak Cluster 10. Thus, if there are 50 BSs in one EC and they have 16 different peak times, these 50 BSs can be consolidated into 16 peak-time clusters.

\begin{figure*}[!t]
\centering
\includegraphics[width=7 in]{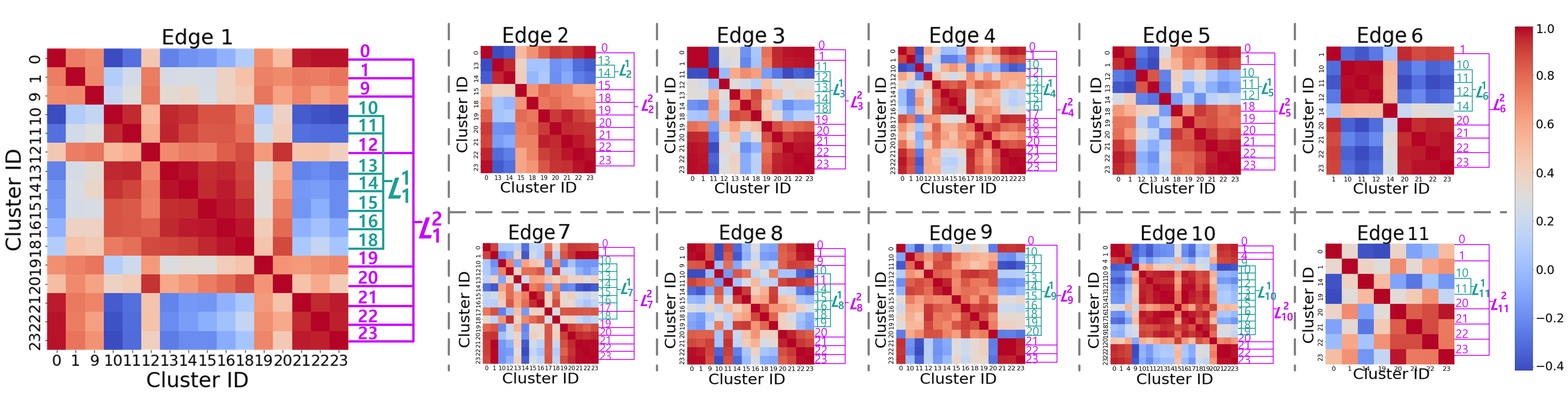}

\caption{Urbanflow Peak Clustering (UPC) for 11 edges, where each edge is divided into peak clusters \( P \) based on peak time patterns on the x- and y-axes. The clusters, in each edge cloud $i$, are consolidated into \( m = 2 \) logical clusters, represented in \textcolor{greenn}{green \( (L_i^1 \))} and \textcolor{pinkk}{pink \( (L_i^2 \))}. Each \( L_i^j \) denotes a logical cluster where \( i \in \{1, \dots, 11\} \) represents the edge ID, and \( j \in \{\textcolor{greenn}{1}, \textcolor{pinkk}{2}\} \) is the logical cluster ID.}
\label{fig_6}
 
\end{figure*}

    As depicted in Fig.~\ref{fig_6}, the initial peak-time clustering step results in several peak clusters $P$ for each of the 11 ECs. For example, $EC_1$ has 16 clusters while $EC_{11}$ has 8.  Each peak-time cluster contains BSs whose traffic peaks at the same hour, capturing the temporal diversity across sites. Therefore, the PCC is applied to merge the correlated $P \in \mathcal{P}$ into \textbf{two} logical clusters, such that the averaged pairwise PCC is maximum, as explained in the consolidation step of Algorithm S1. The selection of $m = 2$ is supported by concrete reasons: (i) Fig.~5’s correlation matrices consistently exhibit two dominant high-correlation groups; (ii) domain studies (e.g., in electricity demand) show that usage patterns typically fall into two distinct peak groups~\cite{navarro2016probabilistic}; and (iii) literature on lightweight hierarchical edge coordination emphasizes minimal yet representative grouping. Oakestra, for instance, consolidates clusters into two tiers to balance accuracy and overhead~\cite{vishwanathan2023oakestra}. The pink and green lines in Fig.~\ref{fig_6} illustrate the consolidation of multiple $P$ into a simplified set of $m = 2$ logical clusters. To summarize, a logical cluster $L_i^j$ represents $EC_i$ and its logical ID $j$ where $i \in \{1,2, \dots 11\}$, and $j \in \{1,2\}$, given $m = 2$. For example, $L_1^1$ in green and $L_1^2$ in pink in Fig. 5 represent two logical clusters in $EC_1$.

\begin{figure}[t]
\centering
\includegraphics[width=3.5in]{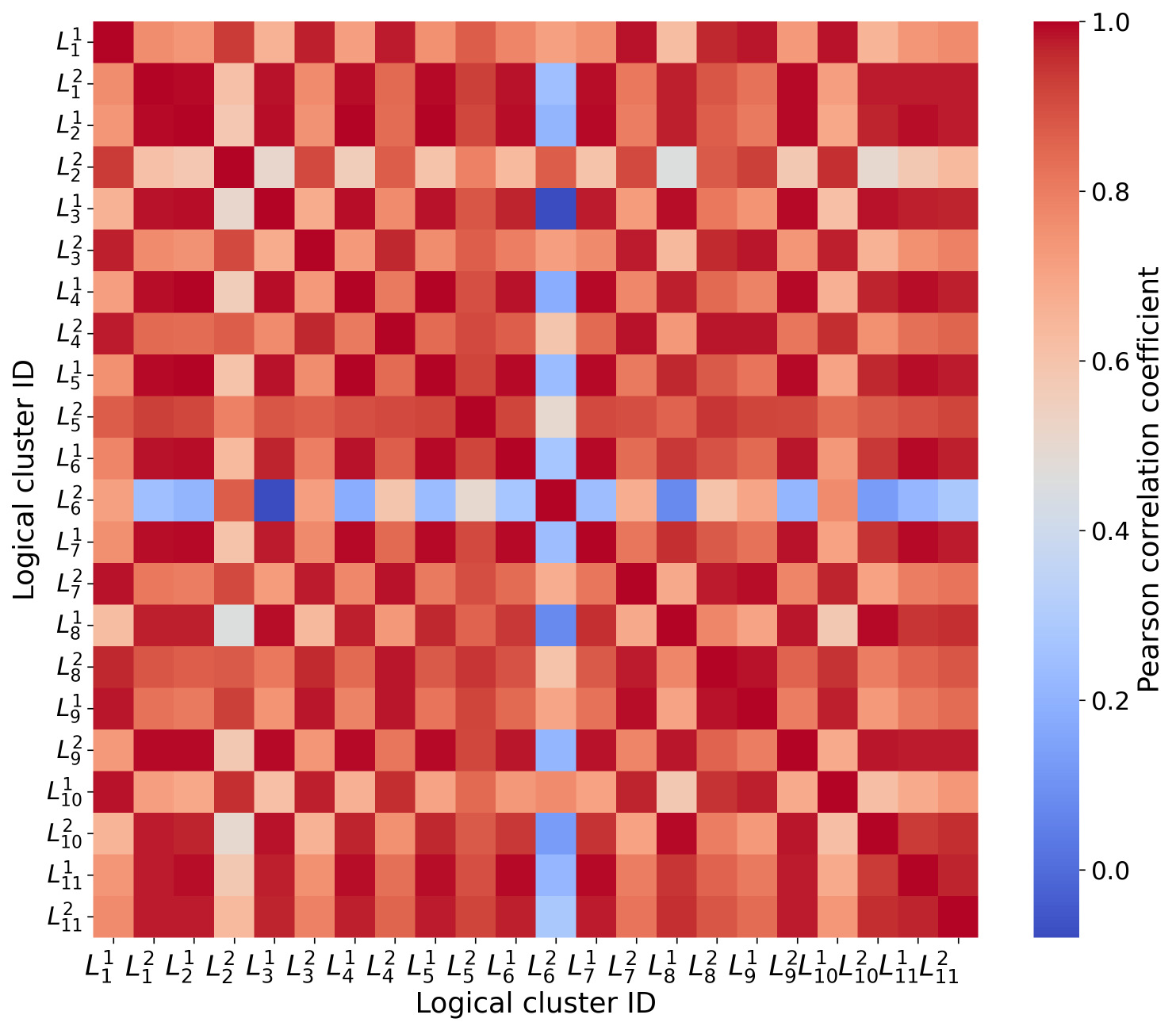}
\caption{Correlation between traffic patterns of all $L \in \mathbb{L}$ from previous Fig. \ref{fig_6}.}
\label{fig_7}
\end{figure}

After forming 22 logical clusters, the average traffic of each cluster is transmitted to the core network, where the correlation matrix between all these logical clusters is calculated, as illustrated in the heatmap Fig. \ref{fig_7}. It is evident that a variety of correlations exist among the logical clusters. As discussed in Section 3, it is advantageous for each logical cluster to collaborate only with those that exhibit a high correlation. For example, $L_1^2$  exhibits a higher correlation with twelve other logical clusters, whereas $L_6^2$ has only $L_2^2$ in its collaboration set. The core network sends the collaboration sets to the respective logical clusters in their ECs. The correlation threshold value for collaboration is set as 0.95 to ensure that the core network forms the collaboration set of each cluster only with those clusters that have the potential to improve the prediction accuracy of the source cluster.

To this end, all logical clusters have been established, and collaboration sets have been received from the core network. Each $L$ now receives data from its respective BSs and forms a six-tuple $\langle$\textit{time}, \textit{edge\_ID}, \textit{BS\_ID}, \textit{coordinates}, \textit{number of users}, \textit{traffic}$\rangle$. Sequences of length $q$ are generated for each BS over time, represented as $\langle s_1, s_2, ..., s_q \rangle$ , where each $s_w$ in the sequence corresponds to the six-tuple at time step $w \in \{1, \dots, q\}$. These sequences are divided into several overlapping input sequences of length $q$. For instance, the first input sequence of length $q = 4$ for any $BS_{i}$ is $\langle s_1, s_2, s_3, s_4 \rangle$, followed by another sequence $\langle s_2, s_3, s_4, s_5 \rangle$, and so on. This approach allows the model to understand the continuous nature of traffic and learn time-series dependencies among the sequences to predict traffic and user counts at $q+1$. The generated sequences are randomly shuffled and split into 70\%, 15\%, and 15\% for training, validation, and testing, respectively.



The traffic and user count predictions are evaluated using MAE, measuring the average absolute error across logical clusters. Training is configured with 1,000 epochs, a batch size of 32, a 0.005 learning rate, and a 0.2 dropout rate to mitigate overfitting. Data is sampled at 60-minute intervals, and the dilated factor, expressed as \(2^i\), expands the model's receptive field.


\setcounter{figure}{6}

\begin{figure}[!t]
\centering
\begin{minipage}{0.45\textwidth}
   \includegraphics[width=\textwidth]{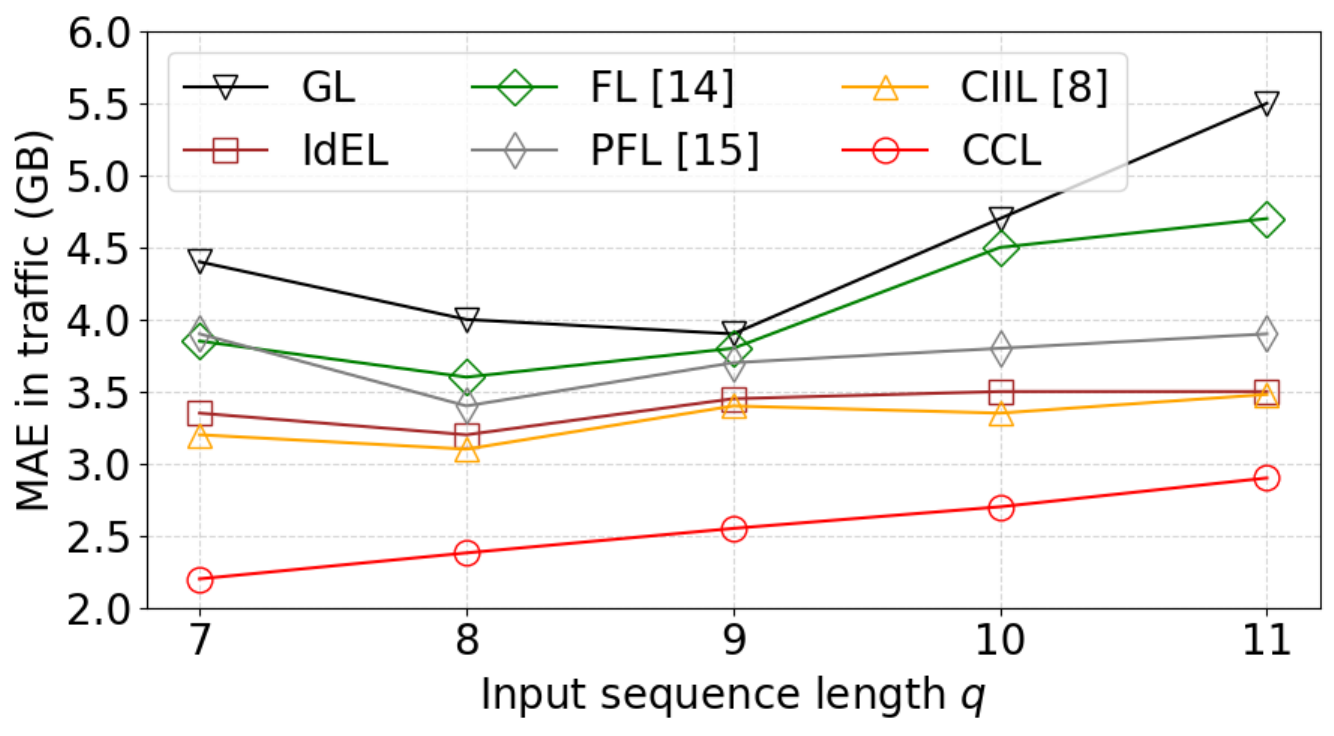}
   \caption*{{(a) Traffic ($\hat{\tau}_{q+1}$) prediction task}}
   \label{fig3a} 
\end{minipage} 
\hfill
\begin{minipage}{0.45\textwidth}
   \includegraphics[width=\textwidth]{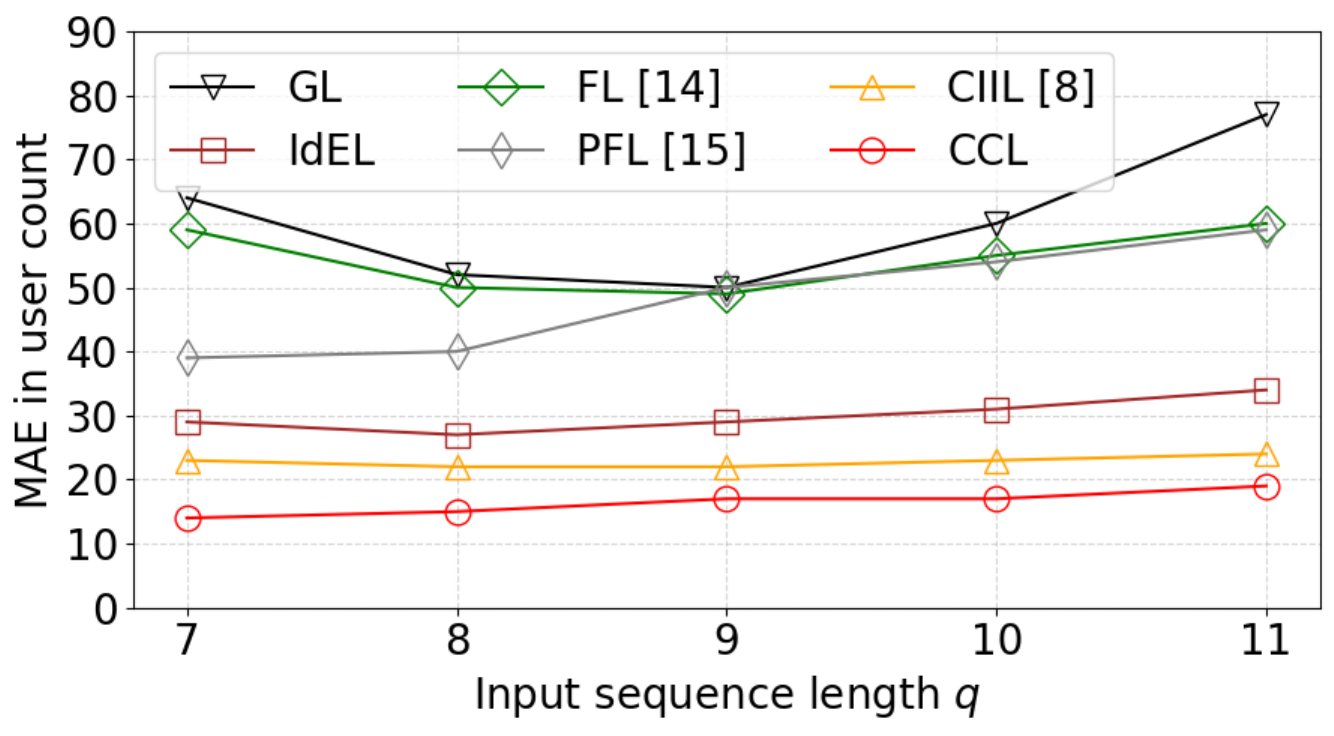}
   \caption*{{(b) User count ($\hat{u}_{q+1}$) prediction task}}
   \label{fig3b}
\end{minipage}
\hfill
    \caption{MAE comparison of CCL with state-of-the-art, averaged across all logical clusters.}
    \label{fig_8}

\end{figure}

\subsection{Experimental Results}

\subsubsection{CCL for Predicting Traffic $\hat{\tau}$ and User Count $\hat{u}$}

Fig.~7(a) compares the MAE performance of CCL and five benchmark strategies for predicting $\tau_{q+1}$ across varying $q$. The benchmark strategies include GL, IdEL, FL, PFL, and CIIL. GL uses a single shared model across all BSs. IdEL trains each logical cluster independently. FL aggregates model updates across ECs without sharing raw data. PFL further fine-tunes locally, while CIIL passes the model states sequentially across clusters. The global model achieves its lowest MAE of 3.9~GB at $q = 9$, corresponding to a 13\% error relative to the dataset’s average BS traffic of 30~GB per hour, as referenced in Fig.~4. IdEL achieves its lowest MAE of 3.22~GB at $q = 8$, reducing the error to 10.7\% by avoiding interference from contrasting traffic patterns, although the lack of inter-cluster knowledge sharing prevents it from generalizing effectively. FL achieves its lowest MAE of 3.6~GB at $q = 8$, corresponding to a 12\% error, though it still suffers from uncorrelated BSs influencing the learning process. PFL improves upon FL with a minimum MAE of 3.7~GB at $q = 9$, benefiting from local fine-tuning after global updates. CIIL reaches its lowest MAE of 3.19~GB at $q = 8$, reducing the error to 10.6\%, since sequential updates retain localized traffic behaviors while still propagating cluster-level knowledge. CCL outperforms all approaches, achieving the lowest MAE of 2.19~GB at $q = 7$, corresponding to a 7.3\% error, as its cross-cluster learning explicitly aligns peak traffic patterns across edges, enabling it to capture both local and global variations more effectively than the other schemes.

The task of user count prediction $\hat{u}$ in Fig. 7(b) follows trends similar to traffic prediction. The GL approach achieves its lowest MAE of 50 at $q = 9$, corresponding to an 11.2\% error relative to the dataset’s average user count of 445 per BS (see Fig. \ref{fig_5}). IdEL improves upon this, achieving its lowest MAE of 27 at $q = 8$, reducing the error to 6.1\% by learning localized patterns. FL reaches its minimum MAE of 49 at $q = 9$, corresponding to an 11\% error, while PFL further refines FL’s performance, achieving an MAE of 39 at $q = 7$, yielding an 8.8\% error. CIIL significantly enhances prediction accuracy, reaching its lowest MAE of 22 at $q = 9$, representing a 4.9\% error. CCL outperforms all approaches, achieving the lowest MAE of 14 at $q = 7$, corresponding to a 3.1\% error. These results underscore the advantage of CCL in ensuring collaboration only among highly correlated logical clusters.

\subsubsection{DUPS}

Once CCL predicts the traffic and user count, the DUPS utilizes these values to pool the DU servers. To demonstrate the versatility of the DUPS across diverse traffic environments, three edges from our dataset are deliberately selected based on their traffic characteristics. The first selected edge $EC_{10}$ exhibits the highest traffic in the dataset, averaging 4.5 TB per hour. The second edge, $EC_{3}$, represents a moderate-traffic environment with an average of 3.5 TB per hour. The final edge, $EC_{11}$, represents a low-traffic scenario, averaging 1.89 TB per hour. This data-driven selection enables the comparative evaluation of DUPS in high-, moderate-, and low-traffic scenarios, showcasing its adaptability across varying network conditions. Several experiments have been conducted to determine the optimal configuration parameters for the DRL environment in all three ECs, aimed at reducing the number of DU servers while maintaining the FH latency requirements. These parameters include $\alpha$, $\beta$ used in the reward function, and DU server capacity $d$, and their selection for all three edges is provided as ablation studies in the next subsection.

\begin{figure*}[htbp]
\setcounter{figure}{7}

\centering   
        \includegraphics[width=\linewidth]{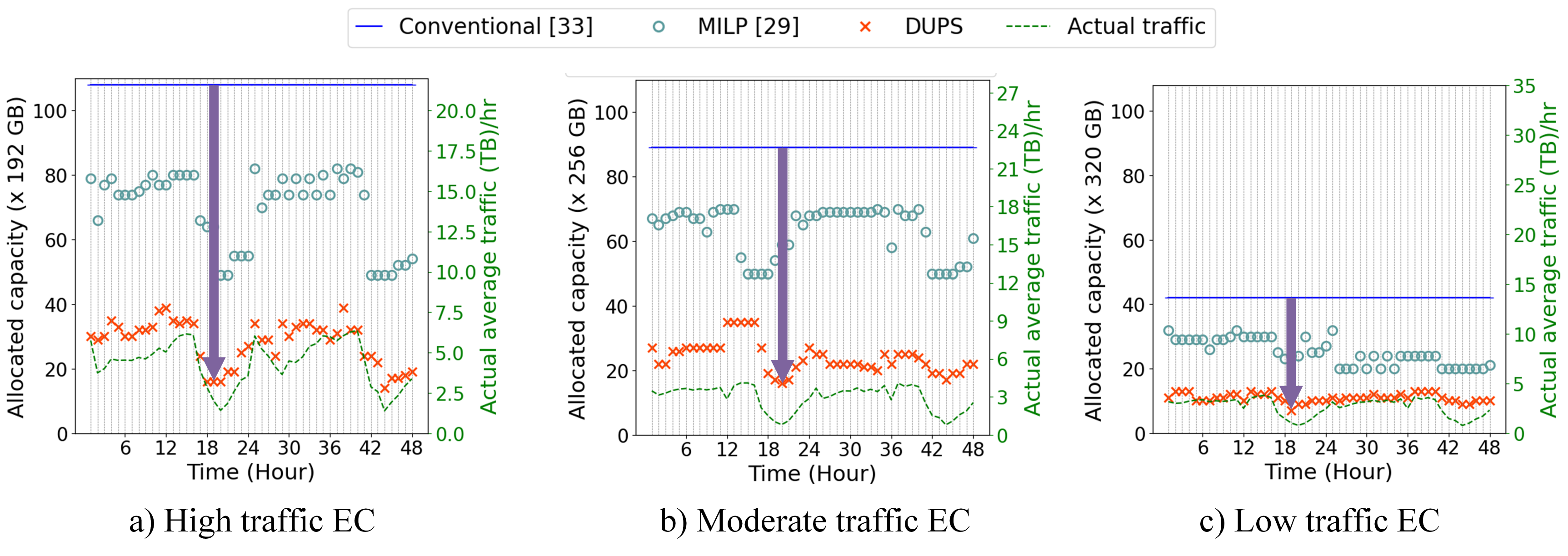}
   \caption{Number of DU servers set by Conventional, MILP, and DUPS over two days under contrastive network traffic edges. The purple arrow highlights the reduction by DUPS compared to the Conventional approach.}
   \label{fig_13} 
\end{figure*}

DUPS is compared with conventional and state-of-the-art MILP strategies using the respective optimal values of $\alpha$, $\beta$, and $d$ for each of the three edges: $EC_{10}$, $EC_{3}$, $EC_{11}$. In the conventional approach, each BS keeps its DU server in an always-active state. Fig. \ref{fig_13} illustrates this comparison, where the green dotted line represents the actual traffic of the respective ECs, the blue solid line shows the DU servers that are always active in the conventional approach \cite{r35a}, the brown line indicates the DU servers proposed by MILP \cite{r39}, and the red line represents those managed by DUPS. For the high-traffic edge $EC_{10}$, DUPS achieves a significant reduction in the number of DU servers, particularly during low-traffic periods. During these periods, DUPS reduces the required DU servers by 89\% and 81\% compared to the conventional approach and MILP, respectively. This significant reduction is attributed to the efficient resource allocation by DUPS. For the moderate-traffic edge $EC_{3}$, DUPS reduces the number of DU servers by 86\% and 77\% compared to the conventional method and MILP approach, respectively. Finally, for the low-traffic edge $EC_{11}$, DUPS reduces the DU server count by 78\% and 52\% compared to the conventional approach and MILP, respectively. These reductions highlight the effectiveness of DUPS in resource management, delivering optimal performance and cost efficiency under varying network conditions.

\begin{table*}[t]
\centering
\caption{Comparison of conventional \cite{r39}, MILP \cite{r35a}, and DUPS in terms of average DU servers, power consumption, energy efficiency, and cost savings across high-, moderate-, and low-traffic ECs. Power consumed per server is 650W \cite{r46} and cost per kWh $r$ is USD $0.13$ \cite{r49}.}
\resizebox{\textwidth}{!}{
\begin{tabular}{|c|c|c|c|c|c|c|c|c|c|}
\hline
\multirow{3}{*}{\textbf{Edge}} & \multicolumn{3}{c|}{\multirow{2}{*}{\textbf{Average DU servers}}} & \multicolumn{2}{c|}{\textbf{Power consumption reduction ($p$) }} & \multicolumn{2}{c|}{\textbf{Energy efficiency improvement (\%)}} & \multicolumn{2}{c|} {{\textbf{Cost saved (USD/h)}}} \\

 & \multicolumn{3}{c|}{\multirow{1}{*}{\textbf{ }}} & \multicolumn{2}{c|}{\textbf{in Watts by DUPS compared to}} & \multicolumn{2}{c|}{\textbf{by DUPS compared to}} & \multicolumn{2}{c|} {{\textbf{by DUPS compared to}}} \\ \cline{2-10}

\rule{0pt}{10pt} \textbf{traffic} & \textbf{Conv.} & \textbf{MILP} & \textbf{DUPS} & $\textbf{Conv.}$ & $\textbf{MILP}$ & \textbf{Conv.} & \textbf{MILP}  & \textbf{   Conv.} & \textbf{MILP} \\ 
 
& $n_{\text{conv.}} $ & $n_{\text{MILP}}$ & $n_{\text{DUPS}}$ &  $(n_{\text{conv.}} - n_{\text{DUPS}}) \times 650$ & $(n_{\text{MILP}} - n_{\text{DUPS}}) \times 650$ & $\frac{n_{\text{conv.}} - n_{\text{DUPS}}}{n_{\text{conv.}}} \times 100$ & $\frac{n_{\text{MILP}} - n_{\text{DUPS}}}{n_{\text{MILP}}} \times 100$ &  $\frac{p \times r}{1000}$ &  $\frac{p \times r}{1000}$\\

\hline
\rule{0pt}{11pt} \textbf{High} & 108 & 70 & 27 & 52650 & 44823 & 75 & 61.4 & 6844 & 5827 \\ 
\hline
\rule{0pt}{11pt} \textbf{Moderate} & 89 & 63 & 24 & 42250 & 40276 & 74.1 & 57.1 & 5492 & 5235 \\ 
\hline
\rule{0pt}{11pt} \textbf{Low} & 42 & 25 & 11 & 20150 & 15589 & 73.8 & 56.0 & 2619 & 2026 \\ 
\hline
\end{tabular}}
\label{tab:energy_efficiency_comparison}
\end{table*}

Beyond its exceptional performance during low-traffic periods, DUPS is also evaluated for its effectiveness over the entire operational timeline. DUPS significantly improves energy efficiency and cost savings by reducing the number of active DU servers at any given time, as provided in Table \ref{tab:energy_efficiency_comparison}. A single DU server consumes approximately 650 Watts \cite{r46}, equivalent to 0.65 kWh per hour. Thus, the total energy consumption for the conventional strategy is \( 0.65\) times the number of active DU servers. For the high-traffic edge $EC_{10}$, the conventional strategy requires 108 DU servers, which MILP reduces to an average of 70 servers, and DUPS further lowers to an average of 27 servers. In other words, DUPS achieves a 75\% reduction in both the average number of servers and energy consumption compared to the conventional strategy and a 61.4\% reduction compared to MILP. For the moderate-traffic edge, DUPS reduces the average number of servers and energy consumption by 74.1\% compared to the conventional strategy and by 57.1\% compared to MILP. Similarly, for the low-traffic edge, these reductions are 73.8\% and 56\%, respectively.

\begin{figure}[!t]
\centering   
        \includegraphics[width=0.95\linewidth]{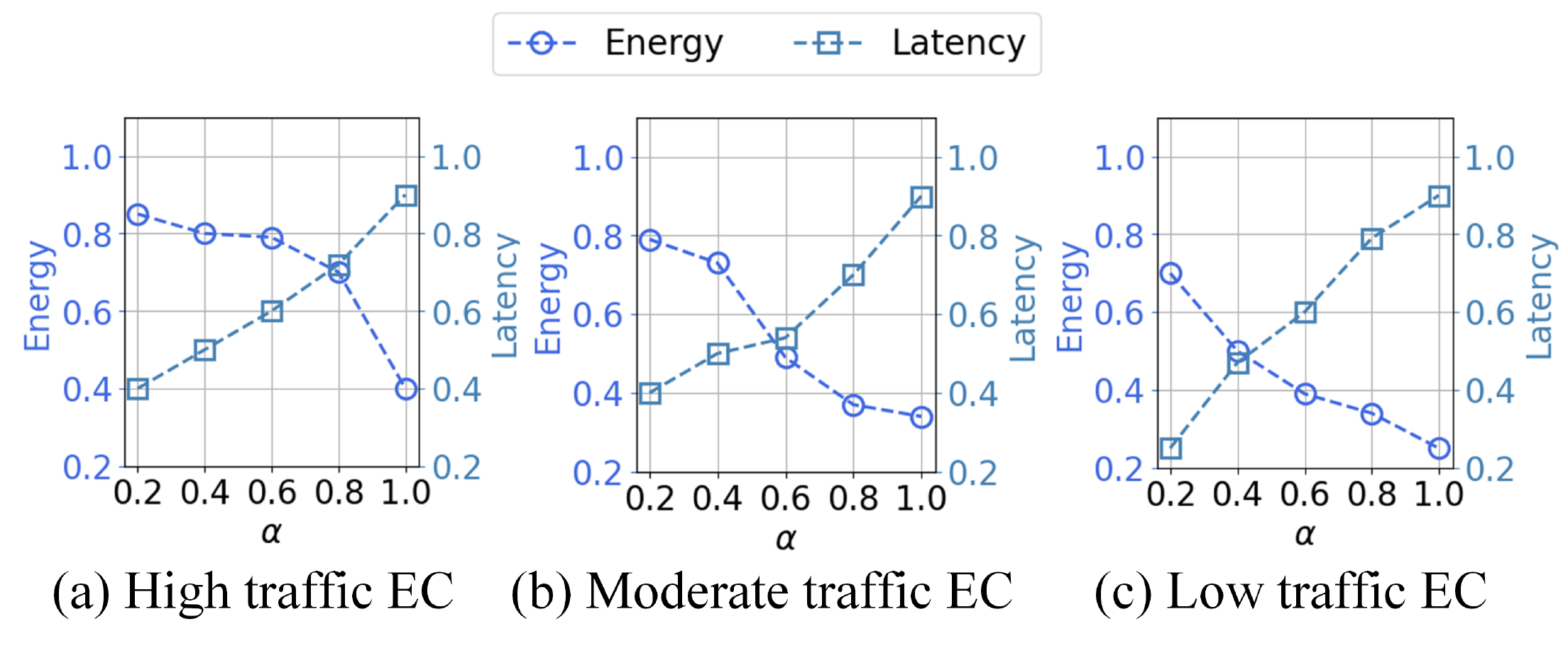}
        \caption{Energy-latency trade-off over varying the coefficient parameter $\alpha$ for contrast traffic-intensive  edges.}
   \label{fig_10} 
\end{figure}

\begin{figure}[!t]
\centering   
        \includegraphics[width=\linewidth]{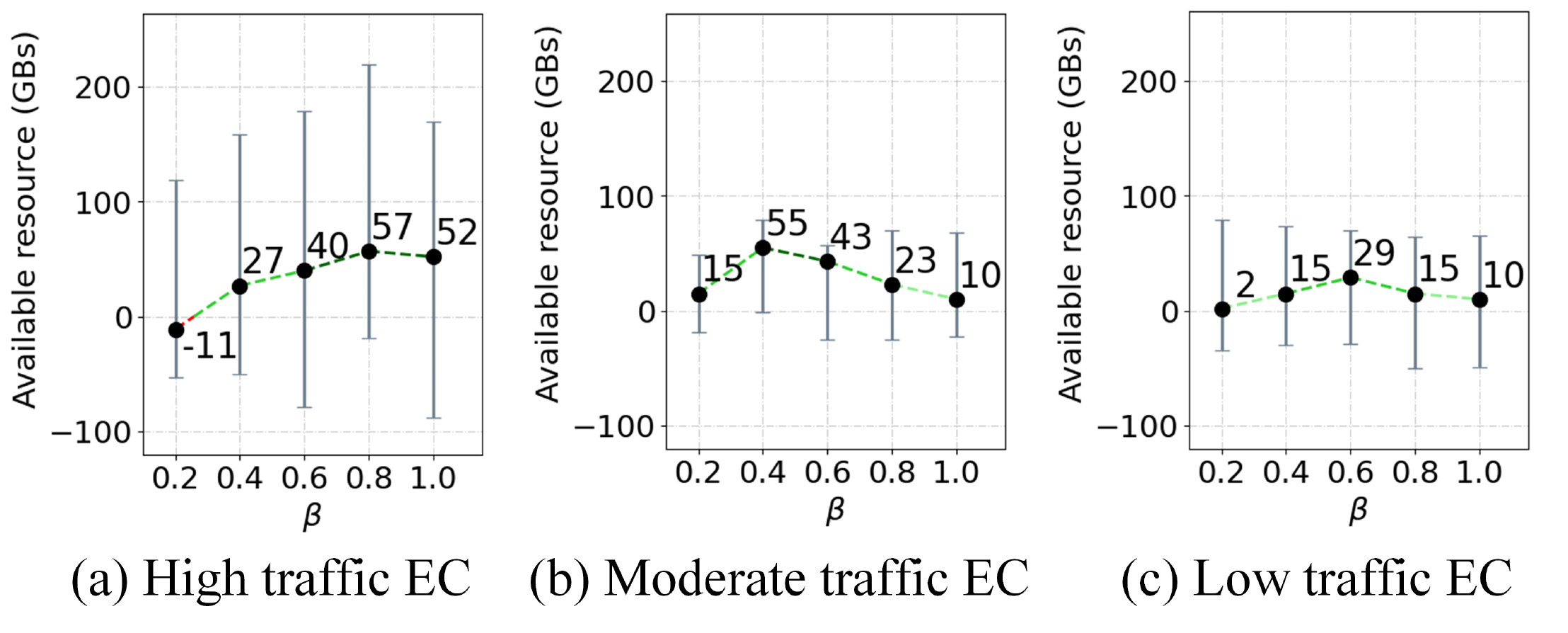}
    \caption{Average available (surplus+/shortage-) memory resource (GB) over varying penalizing parameter $\beta$ for contrastive traffic edges. }
   \label{fig_11} 
\end{figure}

\paragraph{Ablation Studies of DUPS} Fig. \ref{fig_13} presents three edge clouds, $EC_{10}$, $EC_{3}$, and $EC_{11}$, representing high-, moderate-, and low-traffic edges, respectively, to illustrate the effectiveness of the proposed DUPS under varying conditions. Several experiments have been conducted to determine the optimal configuration parameters for all three DRL environments, aimed at reducing the number of DU servers while maintaining FH latency requirements. These parameters include $\alpha$, $\beta$ used in the reward function, and the DU server capacity $d$.

Selecting an optimal value of \( \alpha \) is crucial for balancing energy and latency in the reward function, with fixed parameters \( \beta = 0.5 \) and $d$ = 192 GB. Energy and FH latency are normalized between 0 and 1. For energy, 0 represents a DU server being off, while 1 represents 100\% CPU utilization, calculated using the energy model from \cite{r45}. For latency, 0 indicates the DU server is positioned very close to the RU of the BS, and 1 represents the maximum allowable FH latency of \( 75 \, \mu \text{sec} \). Fig. \ref{fig_10} compares these parameters across different \( \alpha \) values. For the high-traffic edges, a higher \( \alpha \) prioritizes energy savings, allowing the system to tolerate greater FH latency. This encourages agents to turn off more servers, even at the cost of redirecting traffic and increasing FH latency. In contrast, in low-traffic regions with sparsely distributed BSs, a lower \( \alpha \) emphasizes latency, discouraging agents from redirecting traffic to distant DU servers, thereby ensuring better performance for the limited traffic.

The next parameter is $\beta$, which controls the penalty given to agents if they violate either the FH latency bound or redirect traffic to an already overloaded DU server, causing the network traffic to be rejected. Fig. \ref{fig_11} illustrates the effect of varying $\beta$ on the total capacity of DU servers offered by the DRL. For instance, if the DRL offers 15 DU servers, each with a capacity of 192 GB, the total capacity would be $15 \times 192$ = 2.8 TB. If the actual traffic for edge $EC_i$ is about 2.9 TB, resulting in a traffic rejection of 0.1 TB. On the other hand, if the actual traffic is 2.2 TB, the DRL would offer a surplus of 0.6 TB. This analysis is conducted with a fixed $\alpha$ of 0.8, 0.6, and 0.4 for the three edges $EC_{10}$, $EC_{3}$, and $EC_{11}$, respectively, while keeping  $d$ = 192 GB. For a high-traffic edge $EC_{10}$, a higher $\beta$ is crucial because high-traffic edges are more likely to violate FH latency bounds or overload nearby servers when a DU server is turned off. Assigning a higher penalty discourages the DRL from actions that lead to traffic rejection, ensuring balanced server usage and better performance under heavy traffic loads. Conversely, in low-traffic edges, the risk of exceeding FH latency bounds or overloading servers is minimal due to the lighter load. Therefore, a lower $\beta$ allows the DRL to prioritize energy savings and reduce unnecessary penalties, enabling efficient use of resources without compromising service quality.

\begin{figure}[t]
\centering
\begin{minipage}{0.49\textwidth}
   \includegraphics[width=\textwidth]{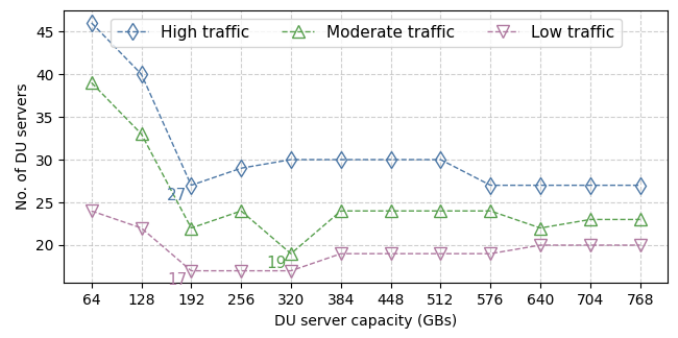}
   \caption*{{(a) Average number of DU servers.}}
   \label{fig3a} 
\end{minipage} 
\hfill
\begin{minipage}{0.49\textwidth}
   \includegraphics[width=\textwidth]{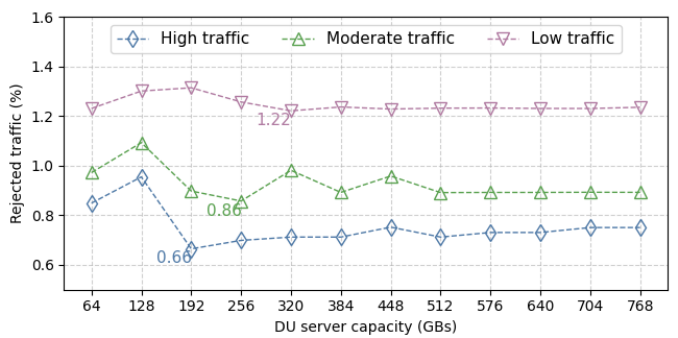}
   \caption*{{(b)  Rejected traffic (\%) of actual traffic.}}
   \label{fig3b}
\end{minipage}
    \caption{Effects on system availability with varying DU server capacity for contrasting traffic edges.}
   \label{fig_12} 

\end{figure}

 Once the appropriate values of \( \alpha \) and \( \beta \) are determined, these parameters are used to identify optimal DU server capacities for each edge. Fig. \ref{fig_12} shows the performance of each edge in terms of the average number of DU servers and the percentage of rejected traffic across varying capacities. The high-traffic edge $EC_{10}$ achieves minimal DU servers and rejected traffic at  $d$ = 192 GB, while the optimal capacities for $EC_{3}$ and $EC_{11}$ are $d$ = 256 GB and 320 GB, respectively. The primary objective is to minimize rejected traffic, as it directly impacts customer satisfaction and the credibility of the operators. Additionally, high-traffic edges benefit from smaller DU capacities to reduce overhead, whereas moderate and low-traffic edges require higher capacities to handle traffic fluctuations effectively.

 \subsection{Complexity Analysis of CCL and DUPS}
\paragraph{CCL}


CCL achieves the highest prediction accuracy by selectively collaborating with expert clusters; however, this improvement comes at a cost of additional computational complexity. All schemes (GL, IdEL, FL, and CCL) share the same core model training complexity. However, CCL has additional complexity due to the selection and monitoring of collaborations among logical clusters. Specifically, the complexity arises from two key steps that are unique to CCL, and their time complexity can be summarized as follows:

\begin{enumerate}[label=\roman*), start=1]

\item Construction of Collaboration set:
According to Algorithm 1, constructing the traffic profile for each cluster requires aggregating the 24-hour traffic vectors of all $n$ base stations, which takes $O(nh)$ time (with $h=24$, this simplifies to $O(n)$). Once the cluster profiles are obtained, computing the PCC for every pair of $L$ clusters involves $\binom{L}{2}$ pairs, each requiring $O(h)$ operations, leading to $O(L^2h)$, or simply $O(L^2)$ since $h$ is constant. After correlations are computed, forming the collaboration set for each cluster by scanning its correlation scores against the other $(L-1)$ clusters requires $O(L)$ time per cluster, resulting in an additional $O(L^2)$ overall. Therefore, the total running time of the algorithm is $O(n + L^2)$.

\item Per-epoch validation-loss monitoring: 
As illustrated in Algorithm \ref{algo2}, for given $L$ logical clusters and a training process of $e$ epochs per cluster, checking whether the rate of change in validation loss falls below the threshold $\psi$ requires only one comparison, i.e., $O(1)$, for each epoch. Thus, for a single cluster across all epochs, this costs $O(e)$. For all $L$ clusters, the overall complexity is $O(Le)$. 

\end{enumerate}

\paragraph{DUPS}
As demonstrated in Algorithm \ref{algo3}, there are $n$ number of DUs, $T$ the number of timesteps per episode, $e$ the total number of training episodes, $B$ the replay-buffer capacity, $|\theta_{\text{critic}}|$ the critic parameter count, $|\theta_{\text{actor}}|$ the actor parameter count, and $s$ the total number of training samples. At each timestep, every DU selects an ON/OFF action in $O(n)$ time, and any DU turned OFF may scan up to $(n-1)$ other DUs to redirect traffic; this yields $O(n^2)$ per timestep, so trajectory collection over $T$ timesteps costs $O(T \cdot n^2)$. After each episode, the critic update computes multi-step returns from the buffer and backpropagates through the critic, which costs $O(T \cdot (B + |\theta_{\text{critic}}|))$. The actor update computes advantages and gradients for all $n$ agents, costing $O(T \cdot n \cdot |\theta_{\text{actor}}|)$. Summing these terms and scaling by $e$ episodes, the total runtime is given as:

\begin{equation}
O\bigl(e \cdot T \cdot (n^2 + B + |\theta_{\text{critic}}| + n \cdot |\theta_{\text{actor}}|)\bigr).
\end{equation}

The $O(n)$ action-selection pass per timestep is dominated by the $O(n^2)$ redirection in the worst case and is therefore absorbed into the overall complexity. The framework remains within polynomial-time complexity while enabling adaptive, DRL-driven resource management.

\section{Conclusion and Future Works}

This study addresses the high energy consumption in 5G RAN, which accounts for 50\% of total power usage, highlighting the untapped potential of AI and deep learning to reduce OpEx while ensuring service quality. This paper presents DUPS, a solution that enhances energy efficiency by up to 89\% and 75\% compared to conventional methods and MILP, respectively, by turning off DU servers and redirecting their traffic. To achieve this, a DRL agent is assigned to each BS, making decisions to turn the server on or off. A single critic in the NWDAF calculates the reward and shares it back to each agent based on its contribution. This also includes a penalty if the DRL agent has forcefully turned off the DU server to save energy by violating latency requirements or due to insufficient capacity in the pool to which traffic is redirected. Thus, a single policy is learned by the system based on the given traffic.

The substantial improvement offered by DUPS relies on accurate traffic prediction, achieved through the proposed CCL, which enhances prediction accuracy by enabling logical clusters to collaborate intelligently based on traffic correlation, sharing partial weights at optimal times. This enables logical clusters to effectively learn the missing information from one another. CCL employs MSTNet, which learns the spatio-temporal patterns from data. CCL is validated on a real cellular dataset from one of the biggest cities of Pakistan for traffic prediction, where it outperforms GL, FL, PFL, and CIIL methods by 43.9\%, 39.1\%, 40.8\%, and 31.35\%, respectively. A detailed evaluation shows that the proposed collaborative strategy delivers the best results in the tested settings, making DUPS with CCL essential for improving efficiency and optimizing network performance.


 It is worth noting that while this work focuses on DU pooling in B5G RAN, the same prediction-driven control loop can also support MEC and Digital Twin (DT) services. In MEC, predicted traffic and user counts can be converted to expected task arrivals, allowing the edge to turn compute resources on or off proactively and schedule uplink transmission for delay-sensitive offloading \cite{r43_new}. The same forecasts can indicate when to consolidate or expand cloudlet resources in distributed offloading, reducing system-wide execution cost \cite{r44_new}. In DT and human DT deployments, twin update and inference workloads fluctuate over time, so traffic-aware pooling can be combined with twin placement and migration to cut server energy while meeting latency and synchronization targets \cite{r47_new}. These are promising extensions to explore for future work in practical MEC and DT systems.

\section*{Acknowledgements}
This work was partly supported by the Korea government (MSIT), IITP, Korea, under the ICT Creative Consilience program (IITP-2025-RS-2020-II201821, 30\%); the Development of 6G Network Integrated Intelligence Plane Technologies (IITP-2025-RS-2024-00392332, 30\%); and by the National Research Foundation of Korea (RS-2024-00343255, 40\%).

\section*{Data and code availability}
The complete dataset is private and cannot be publicly shared. However, a sample from one edge cloud is provided solely for research reference, along with the implementation code for all learning schemes and DUPS, at \url{https://github.com/Sardar-Jaffar-Ali/CCL-DUPS-EdgeAnalytics}.

    \bibliographystyle{IEEEtran}
    \bibliography{reference}





\end{document}